\label{} 
\documentclass[prc,twocolumn,showpacs,floatfix,nofootinbib,preprintnumbers,superscriptaddress,amsmath,amssymb]{revtex4-1}

\usepackage{amsmath}           
\usepackage{amsfonts}
\usepackage{graphicx}          
\usepackage{dcolumn}           
\usepackage{bm}
\usepackage[retainorgcmds]{IEEEtrantools}
\usepackage{color}
\usepackage{multirow}

\newcommand{\be}{\begin{equation}}
\newcommand{\ee}{\end{equation}}
\newcommand{\bn}{\begin{eqnarray}}
\newcommand{\en}{\end{eqnarray}}
\newcommand{\ba}{\begin{array}}
\newcommand{\ea}{\end{array}}
\newcommand{\bc}{\begin{center}}
\newcommand{\ec}{\end{center}}
\newcommand{\bml}{\begin{mathletters}}
\newcommand{\eml}{\end{mathletters}}

\newcommand{\gras}[1]{\boldsymbol{#1}}

\begin{document}


\title{Asymptotic normalization coefficients and continuum coupling in mirror nuclei}

\author
{J. Oko{\l}owicz}
\affiliation{Institute of Nuclear Physics, Polish Academy of Sciences, Radzikowskiego 152, PL-31342 Krak\'ow, Poland
}%

\author{N. Michel}
\affiliation{Department of Physics, Post Office Box 35 (YFL), University of Jyv{\"a}skyl{\"a}, FI-40014 Jyv{\"a}skyl{\"a}, Finland}%
\affiliation{
Department of Physics and Astronomy, University of Tennessee, Knoxville, Tennessee 37996, USA
}%

\author{W. Nazarewicz}
\affiliation{
Department of Physics and Astronomy, University of Tennessee, Knoxville, Tennessee 37996, USA
}%
\affiliation{
Physics Division, Oak Ridge National Laboratory, Oak Ridge, Tennessee 37831, USA
}%
\affiliation{
Institute of Theoretical Physics, University of Warsaw, ul. Ho\.za 69,
PL-00-681 Warsaw, Poland 
}%

\author{M. P{\l}oszajczak}
\affiliation{
Grand Acc\'el\'erateur National d'Ions Lourds (GANIL), CEA/DSM - CNRS/IN2P3,
BP 55027, F-14076 Caen Cedex, France
}%

\date{\today}

\begin{abstract}
\begin{description} 
\item[Background] An asymptotic normalization coefficient (ANC) characterizes the asymptotic form of a one-nucleon overlap integral 
required for description of  nucleon-removal reactions.
\item[Purpose] We investigate the impact of the particle continuum on proton and neutron ANCs  for mirror systems from  $p$- and $sd$-shell regions.
\item[Method]  We use  the real-energy and complex-energy continuum shell model approaches.
\item[Results] We studied the general structure of the single-particle ANCs as a function of the binding energy and orbital angular momentum. We computed ANCs in mirror nuclei for different physical situations, including capture reactions to weakly-bound and unbound states. 
\item[Conclusions] 
We demonstrated that the single-particle ANCs exhibit generic behavior that is different for charged and neutral particles.
We verified the previously proposed  relation \cite{timo03, *Timdes05a} between proton and neutron mirror ANCs. We find minor modifications if the spectroscopic strength is either localized in a single state or broadly distributed. For cases when several states  couple strongly to the decay channel, these modifications may reach 30\%.
\end{description}

\end{abstract}


\bigskip

\maketitle

\section{Introduction}
\label{sec:intro}

The ANC method \cite{Blo77, *Loc78, *Blo93} has proved useful as an indirect tool to determine direct capture reaction rates \cite{Xu94, Gag95, *Ros95, *Muk99, Muk07} both to well-bound and weakly-bound states. The ANC of the virtual proton decay of a nucleus $a\rightarrow b+p$ is related to the astrophysical $S$-factor for a proton capture reaction $b(p,\gamma)a$ at stellar energies, and can be obtained in transfer reactions that offer higher cross-sections; hence, it provides an alternative to direct reactions at relevant energies in astrophysical processes \cite{Xu94,Muk07,Nol11}. The method depends on the peripheral nature of low-energy capture reactions, in which case the cross-section is determined by the tail of the radial overlap integral between the wave functions of the final nucleus and initial colliding systems. 

The ANC, or the nuclear vertex constant, for the decay $a\rightarrow b+c$ is proportional to the asymptotic behavior of the  wave function representing the relative motion of particles $b$ and $c$. This quantity is closely related to the   reaction amplitude \cite{Muk07} and -- unlike  the spectroscopic factor -- is invariant under finite-range unitary transformations of  the nucleon-nucleon (NN) interaction \cite{Muk10}; hence is less dependent on the choice of a short-range potential.

Recently, it has been advocated  \cite{timo03, *Timdes05a} that the charge symmetry of the nuclear force could be used to relate the ANC of  the proton decay to the virtual neutron decay of the mirror nucleus. This observation opens an attractive possibility to learn about the decay width of hardly accessible states in proton-rich nuclei from transfer reaction studies in  mirror-bound systems using stable beams. 
It remains, however, an important question to what extent the charge symmetry argument  is sufficient to extract reliable information about the width of an unbound state of some nucleus from the ANC in its bound mirror partner. It is well-known, for example, that threshold effects lead to striking differences in the energy spectra of mirror nuclei having different particle emission thresholds \cite{Ehr51, *Tho52}. Indeed, for near-threshold states, the configuration mixing involving scattering states  strongly depends on the positions of particle emission thresholds in mirror systems (the binding energy effect) \cite{Oko08}, and on different asymptotic behavior of  neutron and proton wave functions. The latter effect leads to the universal behavior of cross sections \cite{Wigner,Breit} and overlap integrals  \cite{my07,Mic07,Mic10}  in the vicinity of  the reaction threshold.

The main objective of this study  is to verify the conjecture of Refs. \cite{timo98, *timo03a, timo03, *Timdes05a, Timdes05, Timdes07,Tit11} that the ratio ${\cal R}$ of ANCs for mirror pairs is both approach-independent and interaction-independent. To investigate  the proposed link  between proton and neutron mirror ANCs  \cite{timo03, *Timdes05a},  we employ  the framework of the nuclear Shell Model (SM) for open quantum systems (OQS), i.e.,  the  continuum shell model (CSM) \cite{oko,micrev}, which  offers a realistic treatment of the configuration mixing  in well-bound, weakly-bound, and unbound states.

The paper is organized as follows. In Sec.~\ref{sec:ANC} we discuss basic features of  ANCs. In particular, we review  properties of  ANCs for proton wave functions in bound nuclei and the limiting behavior of  ANCs  for charged particles and neutrons. The
CSM results for the ANC in mirror $p$- and $(sd)$-shell nuclei are discussed in Sec.~\ref{sec:CSM}. We employ the complex-energy Gamow Shell Model (GSM) and
the real-energy   Shell Model Embedded in the Continuum (SMEC) to describe ANCs for  bound states and resonances. We also analyze the dependence of the ANCs in mirror systems, and their ratios, on the strength of the continuum coupling and configuration mixing. 
Finally,  Sec.~\ref{sec:Concl}  summarizes the  results of our work.

\section{Basic features of the asymptotic normalization coefficient}
\label{sec:ANC}

Let us consider the radiative capture reaction $b+c \rightarrow a + \gamma$, and define the radial overlap function 
$I^a_{bc}$ for a process $ a\rightarrow b+c$. In the asymptotic region,  $I^a_{bc}$ can be written as:
\begin{eqnarray}
I^a_{bc;\ell j}(r) \sim \frac{1}{r}C_{\ell j}W_{-\eta,\ell+1/2}(2\kappa r),
\label{eq1}
\end{eqnarray}
where $W_{-\eta,\ell+1/2}$ is the Whittaker function and $C_{\ell j}$ is the ANC, a quantity characterizing the virtual decay of a nucleus into two particles $b$ and $c$. In Eq.~(\ref{eq1}), $r$ is the relative distance between  $b$ and $c$, 
$\kappa=\sqrt{2\mu S_c^{(a)}/\hbar^2}$ where $S_c^{(a)}$ is the separation energy of particle $c$ in the nucleus $a$, and $\eta=Z_bZ_ce^2\mu/\hbar^2\kappa$ where $\mu$ is the reduced mass of $b+c$. The quantum numbers $\ell$ and $j$ are the orbital angular momentum and the channel angular momentum, respectively. 

In  this paper we shall assume that the particle $c$ is a nucleon (proton or neutron); hence, $S_{n,p}^{(a)}$ is the one-nucleon separation energy. The corresponding radial overlap integral can be written as:
\begin{eqnarray}
I_{bc;\ell j}^a(r)={1 \over \sqrt{2J_a+1}}\sum_{\cal B}
\langle{\Psi}_a^{J_a}||a_{\ell j}^{\dag}({\cal B})||\Psi_b^{J_b}\rangle\langle r\ell j|u_{\cal B}\rangle,
\label{eqex}
\end{eqnarray}
where  $a^+_{\ell j} (\mathcal{B})$ is a creation operator associated with the 
single-particle (s.p.) basis state $|u_\mathcal{B}\rangle$ and $\langle r\ell j|u_{\cal B}\rangle$ is the radial s.p.~wave function. The sum in (\ref{eqex}) runs over the complete s.p.\ basis. The squared norm of the radial overlap integral (\ref{eqex}) defines  the spectroscopic factor $S_{\ell j}$.

In the general case of multi-channel coupling,  the ANC for a radiative capture reaction is defined in terms of  the Hermitian norm $|C|$ of all the contributions corresponding to different couplings of the target state and the state in a parent nucleus:
\begin{equation}
\left|C\right|=\sqrt{\sum_{\ell,j} |C_{\ell j}|^2}.
\label{Cnorm}
\end{equation}

For bound states, $I^a_{bc;\ell j}$ can be well approximated by the product of the spectroscopic amplitude $S_{\ell j}^{1/2}$ and the s.p.\ radial wave function $u_{\ell j}/r$ at a s.p.~energy $-S_c^{(a)}$:
\begin{eqnarray}
I^a_{bc;\ell j}(r) \sim \frac{1}{r}S_{\ell j}^{1/2}u_{\ell j}(r).
\label{eq2}
\end{eqnarray}
For $r\gg R$, where $R$ is the nuclear radius,  $u_{\ell j}$ is given by its asymptotic form
\begin{equation}\label{defbeta}
u_{\ell j}(r)=\beta_{\ell j}W_{-\eta,\ell+1/2}(2\kappa r),
\end{equation}
where $\beta_{\ell j}=\beta_{\ell j}(S_c^{(a)})$  is the single-particle  ANC (SPANC).
Therefore, far from the region of nuclear interaction, $I^a_{bc;\ell j}$ behaves as:
\begin{equation}
I^a_{bc;\ell j}(r) \sim \frac{1}{r}S_{\ell j}^{1/2}\beta_{\ell j}W_{-\eta,\ell+1/2}(2\kappa r).
\label{eq3}
\end{equation}
Hence, $\beta_{\ell j}$ is directly related to the ANC \cite{Muknun05, *Pang07}:
\begin{eqnarray}
C_{\ell j}=S_{\ell j}^{1/2}\beta_{\ell j}.
\label{eq4}
\end{eqnarray}
The relations (\ref{eq2})-(\ref{eq4})  also hold for many-body resonances. Indeed, as  demonstrated in GSM studies \cite{my07},  the overlap function $I^a_{bc;\ell j}$ is well approximated by the product of the spectroscopic amplitude 
$S_{\ell j}^{1/2}$ and the s.p.\ resonance wave function of the average potential, which reproduces the $Q$-value of the  reaction studied.
Let us also recall that the astrophysical ${\cal S}_{bc}$-factor, in the limit 
of zero center-of-mass energy, $E_{\rm cm}=0$,  is simply proportional to  $\beta^2_{\ell j}$ \cite{Xu94}: 
\begin{eqnarray}
{\cal S}_{bc}(0)\sim \beta_{\ell j}^2\equiv \beta_{\ell j}^2(\eta).
\label{eq6}
\end{eqnarray}

\subsection{General  properties of  SPANCs for charged particles}\label{SPANC}

Let $R_f$ ($R_f \gg R$) be the radius of the external region  where the nuclear part of the potential is practically zero. In this region ($r \ge R_f$) the s.p.~wave function
$u_{\ell j}$ is given by Eq.~(\ref{defbeta}), i.e., it is given by $W_{-\eta,\ell+1/2}(2 \kappa r)$. For the normalized bound state, this implies that ${\cal N}_{\rm int} + {\cal N}_{\rm ext} =1$, where 
${\cal N}_{\rm int}$ is the norm of the internal part of the wave function, and 
\begin{equation}\label{extN}
{\cal N}_{\rm ext}
=\beta_{\ell j}^2 \int_{R_f}^{+\infty} |W_{-\eta,\ell+1/2}(2 \kappa r)|^2\,dr
\end{equation}
is the norm of the external part.
Provided that the energy dependence of ${\cal N}_{\rm int}$ is weak (which is a reasonable assumption even if separation energy is close to zero), $\beta_{\ell j}$ should strongly depend on the value
of the  integral in Eq.\,(\ref{extN}).

In terms of the outgoing Coulomb wave function $H^+(\ell,\eta_k,k r)$, the  Whittaker function can be written as
\begin{equation}
W_{-\eta,\ell+1/2}(2 \kappa r) = H^+(\ell,\eta_k,k r) e^{i\frac{\pi \eta}{2} + i\frac{\pi \ell}{2} - i\sigma(\ell,\eta)} \label{Hplus_W_relation},
\end{equation}
where $k = i \kappa$, $\eta_k = -i \eta$, and $\sigma(\ell,\eta)$ is the Coulomb phase shift. To discuss the limiting cases, is useful to introduce the complex turning point
\begin{equation}\label{tp}
z_t = -i \left( \eta + \sqrt{\eta^2 + \ell(\ell + 1)} \right)
\end{equation}
at which  $d^2 H^+(\ell,\eta_k,z)/dz^2 = 0$.

\begin{widetext}
According to the standard properties of the Coulomb wave functions  \cite{Abr70,Gra80}, one obtains the asymptotic expressions:
\begin{IEEEeqnarray}{rCll}
W_{-\eta,\ell+1/2}( 2\kappa r) &\simeq&  \left( \frac{\kappa r}{2 \eta} \right)^{1/4} \exp \left( \eta - \eta \ln (\eta) - 2\sqrt{2 \eta \kappa r}\right) &~\mbox{ for } \kappa r \ll |z_t|  \mbox{ , } \eta \rightarrow +\infty, \label{W_proton_small_z} \\
W_{-\eta,\ell+1/2}(2 \kappa r) &\simeq&  \exp \left( -\kappa r - \eta \ln(2 \kappa r) \right) &~\mbox{ for } \kappa r \gg |z_t|. \label{W_large_z} 
\end{IEEEeqnarray}
\end{widetext}

\subsubsection{Near-threshold limit of SPANCs for charged particles}\label{SPANCp1}

Let us first consider the limit (\ref{W_proton_small_z}) of $\eta \rightarrow +\infty$, which corresponds to very small separation energies. Since $\eta=\kappa_0 /\kappa$, where $\kappa_0=Z_bZ_ce^2\mu/\hbar^2$, the asymptotic part of the s.p.~wave function near the particle emission threshold is
\begin{equation}\label{uprot}
u_{\ell j}(r) \simeq \beta_{\ell j}
\left( \frac{\kappa_0 r}{2\eta^2} \right)^{1/4}  \exp \left( \eta - \eta \ln (\eta) - 2\sqrt{2 \kappa_0 r}\right).
\end{equation}
As the external norm (\ref{extN}) must be finite, and 
$\exp[\eta - \eta \ln (\eta)]/\eta^{1/2} \rightarrow 0$ for 
$\eta \rightarrow +\infty$, 
in the limit of very weak binding $\beta_{\ell j}$ must exhibit the
 universal $\eta-$dependence:
\begin{eqnarray}
\tilde{\beta}_{\ell j}(\eta)=N_{\beta}(\ell,j)\eta^{1/2}\exp(\eta\ln\eta-\eta) ,
\label{buni}
\end{eqnarray}
where $N_{\beta}(\ell,j)$ is a prefactor that depends on the structure of the s.p.~state, in particular $\ell$. 
To assess how quickly the  limit (\ref{buni}) is reached, we performed calculations  for the $1s_{1/2}$, $0p_{1/2}$, and $0d_{5/2}$ single-proton states in $^{17}$F. The s.p.~radial wave function $u_{\ell j}$ was calculated using the Woods-Saxon (WS) potential with the following parameters: the strength of the spin-orbit term $V_{\mbox{so}} = 3.68$\,MeV, radius $R_0=3.214$\,fm, and diffuseness $d=0.58$\,fm. We took the Coulomb potential of  a spherical uniform charge distribution  with the  radius $R_0$. For each $\ell$, $j$, the depth of the central potential has been  adjusted to  the proton separation energy in $^{17}$F, which corresponds to a given value of $\eta$. The value of $N_{\beta}(\ell,j)$  can be extracted from the calculated wave function
at $\eta > 100$ and it is 10.786, 2.914, and 0.42 for the  $1s_{1/2}$, $0p_{1/2}$, and $0d_{5/2}$ states, respectively.
%
\begin{figure}[hbt]
\center
\includegraphics[width=0.45\textwidth]{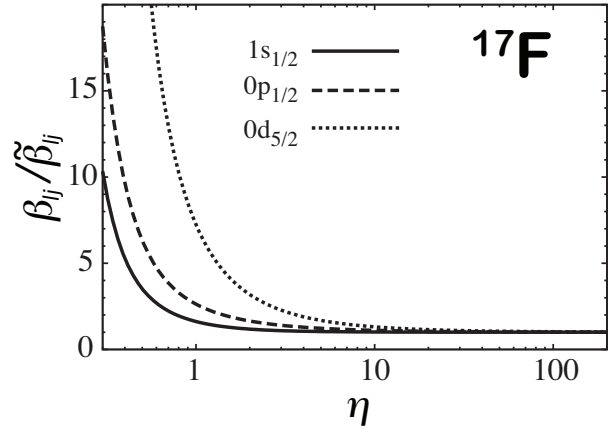}
\caption{\label{fig5} The  $\eta-$dependence of 
${\beta}_{\ell j}/\tilde{\beta}_{\ell j}$
for $1s_{1/2}$, $0p_{1/2}$, and $0d_{5/2}$ s.p.~proton wave functions  in $^{17}$F.}
\end{figure}
Figure~\ref{fig5} shows the $\eta-$dependence of the
ratio ${\beta}_{\ell j}/\tilde{\beta}_{\ell j}$. The asymptotic behavior is reached at  $\eta\approx 1$ for $\ell=0$ and   $\eta\approx 10$ for $\ell=2$ and
the ratio  smoothly decreases with $\eta$ suggesting 
a polynomial dependence on the proton separation energy (or a
Maclaurin series in $1/\eta$). Guided by this result, we write:
\begin{equation}
\label{poly}
\beta_{n \ell j}(\eta)=\eta^{1/2}\exp(\eta\ln\eta-\eta)\sum_{i=0}^\infty \frac{a_i(n,\ell,j)}{\eta^i}.
\end{equation}
It is worth noting that the term $\eta^{1/2}\exp(\eta\ln\eta-\eta)$, governing the rapid growth of SPANC around the threshold, is universal for {\em all charged particles} (proton, deuteron, $\alpha$, $\ldots$), independent of their quantum state. The dependence on the structure is contained in the coefficients $a_i$ of the Maclaurin series, which carry information  on quantum numbers ($n, \ell, j$) of the bound state. 
\begin{table} [htb]
\begin{center}
\caption{Coefficients of the fourth-order Maclaurin series (\ref{poly}) fitted to the
ratio ${\beta}_{\ell j}/\tilde{\beta}_{\ell j}$ of single-proton WS states in $^{17}$F 
in the range of  $S_p <10$ MeV. For the $1s_{1/2}$ state, three values of diffuseness $d$ (in fm)  were used.}
\label{tablo_eta}	
\begin{ruledtabular}
\begin{tabular}{c|ccc|cc}
 & \multicolumn{3}{c|}{$1s_{1/2}$} &  $0p_{1/2}$ & $0d_{5/2}$ \\ [2pt]
$d$ & 0.46 & 0.58 &  0.7  & 0.58 & 0.58 \\ \hline 
$a_0$ & 9.934 & 10.786 & 11.711 & 2.914 & ~0.4200 \\
$a_1$ & ~0.8255 & ~~0.8889 & ~~0.9515 & 3.022 & 1.185 \\
$a_2$ & 5.483 & ~6.393 & ~7.444 & 1.416 &  1.273 \\
$a_3$ & $-0.9821$ & $-1.157$ & $-1.362$ & ~0.1411 & $-0.0408$ \\
$a_4$ & ~0.4598 & ~~0.5635 & ~~0.6945 & ~0.1676 & ~0.2035 
\end{tabular}
\end{ruledtabular}
\end{center}
\end{table}
In the considered example of single-proton states in $^{17}$F, the coefficients $a_i$ were fitted in the range of proton separation energies $S_p^{(a)} <10$\,MeV. 
An excellent fit has been obtained with the first five terms in the expansion (\ref{poly}). The resulting values $a_i$ $(i=0,\ldots,4)$ are listed in Table~\ref{tablo_eta}. To study the model dependence, for $1s_{1/2}$ we considered three values of WS diffuseness $d$. In this case, the coefficients $a_i$ vary by 20-40\% if $d$  changes from the value of 0.46\,fm to 0.7\,fm. 
The variations in $a_i$ with $d$ are further reduced if the r.m.s.~radius of the potential is kept constant while changing $d$. For instance, the changes in $a_0$ are  $\sim1\%$ in the  calculations constrained in such a  way.

\subsubsection{Large binding energy limit of SPANCs for charged particles}\label{SPANCp2}

Now we consider the limit (\ref{W_large_z}) of smaller $\eta$, which corresponds to finite  separation energies that are  small enough so that the $\eta$-variations of ${\cal N}_{\rm int}$ can be neglected.
In this case, the asymptotic part of the wave function shows the usual exponential decay:
\begin{equation}\label{uprot1}
u_{\ell j}(r) \simeq \beta_{\ell j}
\exp \left( -\kappa_0 r/\eta - \eta \ln(2 \kappa_0 r/\eta) \right).
\end{equation}
In this case, in order to keep ${\cal N}_{\rm ext}$ finite, $\beta_{\ell j}$ has to increase   when $\eta$ decreases. Consequently,
when inspecting 
$\beta_{\ell j}$ as a function of $\eta$ one can expect a minimum when $\eta$ changes from small values toward the threshold ($\eta=+\infty$).

\subsubsection{Separation energy dependence of SPANCs for bound proton wave functions}
\label{SPANCp3}

We shall now discuss the behavior of $\beta_{\ell,j}$ in the full energy range.
\begin{figure}[hbt]
\center
\includegraphics[width=0.35\textwidth]{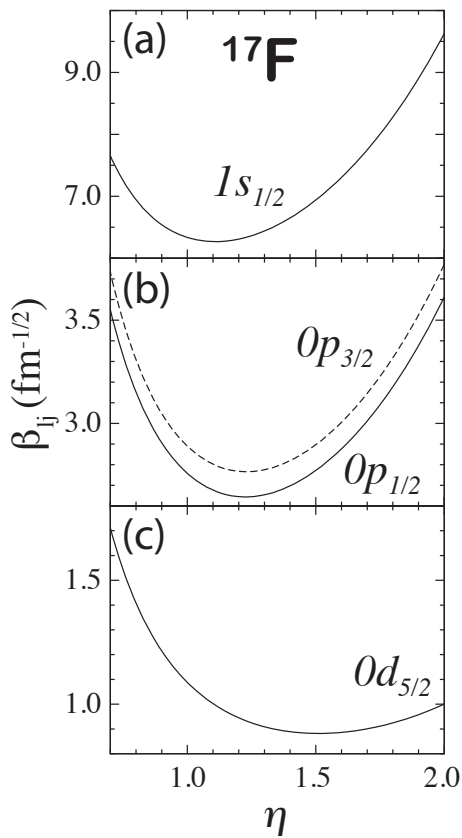}
\caption{\label{fig1}Energy dependence of  $\beta_{\ell j}$ for  the bound-state proton s.p.\ wave functions in $^{17}$F 
with  $1s_{1/2}$ (a); $0p_{1/2}$ and $0p_{3/2}$  (b); and $0d_{5/2}$ (c).}
\end{figure}
Figure~\ref{fig1} shows $\beta_{\ell j}(\eta)$ for the  p + $^{16}$O capture reaction at $E_{\rm cm}=0$ in channels where the proton has a relative angular momentum $\ell=0$ ($1s_{1/2}$), $\ell=1$ ($0p_{1/2}$ and $0p_{3/2}$), or $\ell=2$ ($0d_{5/2}$). We used the  same WS potential as in Sec.~\ref{SPANCp1}.
It is seen that the behavior of SPANC as a function of $\eta$ follows general considerations of Sec.~\ref{SPANCp2}.  Namely,
with increasing $\eta$,  $\beta_{\ell,j}$ first decreases until  a certain minimum value of $\eta=\eta_{\rm crit}$ is reached, and then increases again as the separation energy decreases towards the $S_p^{(a)}=0$ threshold. We may thus conclude that $\eta_{\rm crit}$ separates the regimes of strong 
($\eta < \eta_{\rm crit}$) and weak ($\eta > \eta_{\rm crit}$) binding for a given partial wave and $Z_a$. 
In general, $\eta_{\rm crit}$ scales approximately linearly with $Z_b$ and it strongly depends  on the angular momentum $\ell$ of the proton. On the other hand, the dependence on the channel angular momentum $j$ is  weak. The magnitude of $\beta_{\ell,j}$  decreases with $\ell$. The case of   $\ell=0$ and 
large $\eta$ (and $\beta_{\ell=0,j=1/2}$) shown in Fig.~\ref{fig1}(a) is characteristic  of a   proton  halo. However, as seen in Fig.~\ref{fig1} and discussed in Sec.~\ref{SPANCp2}, the large values of 
$\beta_{\ell,j}$ are also expected for very bound states, so a large SPANC  is not an indicator of a proton halo.

The extreme regimes  of SPANC can be characterized by the complex turning point $z_t$ given by Eq.~(\ref{tp}). 
The low-binding regime of $\beta_{\ell,j}$ is reached for $\kappa r \ll |z_t|$ and  $\eta \rightarrow +\infty$. In this region, characterized by the condition
\begin{equation}
 \left( \eta + \sqrt{\eta^2 + \ell(\ell + 1)} \right)\eta \gg \kappa_0 R_f,
\end{equation}
the proton wave function behaves asymptotically as
$u_{\ell j}(r) \propto r^{1/4}  \exp(- 2\sqrt{2 \kappa_0 r})$, i.e., its decay is slower than exponential. The strong-binding regime is reached at
$\kappa r \gg |z_t|$, i.e.,
\begin{equation}
 \left( \eta + \sqrt{\eta^2 + \ell(\ell + 1)} \right)\eta \ll \kappa_0 R_f.
\end{equation}
Here, the proton wave function shows the expected exponential decay 
(\ref{uprot}).

\subsubsection{Near-threshold behavior of charged particle radiative capture cross sections}
\label{sigmap}

Formally, one can discuss the charged particle radiative capture cross section in the two limits: (i)
$\eta_{CM}\rightarrow \infty$ ($E_{\rm CM}\rightarrow 0$), and (ii) $\eta\rightarrow \infty$ ($S_c^{(a)}\rightarrow 0$). Whereas the CM energy of the system $\langle b+c\rangle$ can be varied experimentally, the charged particle separation energy is fixed for any state of the $a$-nucleus and, therefore, the limiting behavior of the radiative capture cross section when $\eta\rightarrow \infty$ cannot be studied experimentally in a single physical system. 
For a fixed value of $\eta$, the radiative capture cross-section $\sigma_{c\gamma}(\eta_{CM},\eta)$ is exponentially reduced  in the first limit ($\eta_{CM}\rightarrow \infty$) as:
\begin{eqnarray}
\lim_{\eta_{CM}\rightarrow \infty}\sigma_{c\gamma}={\cal S}_{bc}(\eta_{CM},\eta) \frac{\exp(-2\pi\eta_{CM})}{E_{\rm CM}}.
\label{asym1}
\end{eqnarray}

As immediately follows from Eq.~(\ref{eq6}) and discussion in Sec.~\ref{SPANCp1},
in a sub-threshold regime ($\eta\rightarrow \infty$), the radiative capture cross-section $\sigma_{c\gamma}(\eta_{CM},\eta)$
for a fixed value of $\eta_{CM}$ diverges as:
\begin{eqnarray}
\lim_{\eta \rightarrow \infty}\sigma_{c\gamma}\propto{\cal S}_{bc}(\eta_{CM},\eta)\propto \eta \exp(2\eta\ln(\eta) - 2\eta).
\label{asym2}
\end{eqnarray}
The asymptotic behavior of 
$\sigma_{c\gamma}$ given by Eqs.~(\ref{asym1}) and  (\ref{asym2}) is expected  for any charged particle radiative capture reaction.

\subsubsection{Survey of experimental proton  SPANCs}
\label{SPANC-proton-exp}

\begin{figure}[hbt]
\center
\includegraphics[width=0.40\textwidth]{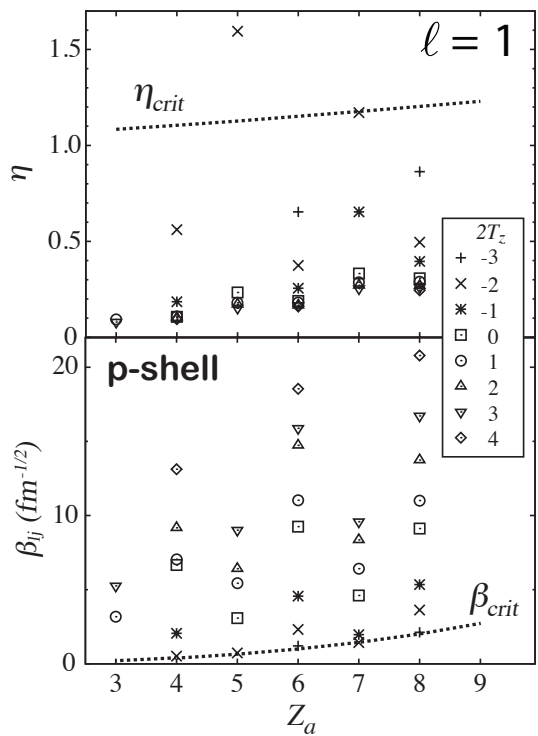}
\caption{\label{fig2} 
Experimental values of $\eta$ (top) and $\beta_{\ell,j}$ (bottom)  for bound proton states with $\ell=1$ and $j=1/2, 3/2$  in  various $p$-shell nuclei ($Z_a, T_z$). The curves of $\eta_{\rm crit}$ and $\beta_{\rm crit}$  are calculated for
$T_z=-1/2$ nuclei. See text for details.}
\end{figure}

\begin{figure}[hbt]
\center
\includegraphics[width=0.40\textwidth]{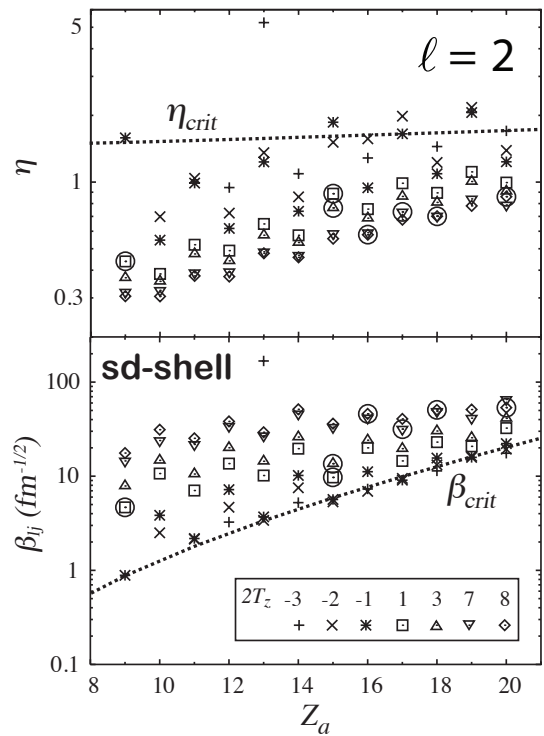}
\caption{\label{fig3}Similar as in Fig.~\ref{fig2} except for $\ell=2$ bound proton wave functions in $sd$-shell nuclei. For certain nuclei, the radiative proton capture  with the channel angular momentum $j=3/2, 5/2$ cannot populate the ground state of a nucleus $a$. In such cases  (encircled symbols), we plot   $\eta$ and $\beta_{\ell,j}$ for the lowest-energy excited state that is populated by the  capture  with $\ell=2$ protons.}
\end{figure}

Figures~\ref{fig2} and \ref{fig3}   survey  the values of $\beta_{\ell,j}$ and $\eta$ for the proton emission channels with $\ell=1$ ($p$-shell) and $\ell=2$ ($sd$-shell), respectively.
The $Z_a-$dependence of $\eta_{\rm crit}$ is fairly weak in both cases.
With the exception of the known proton halo $^8$B, all other $p$-shell nuclei belong to the class $\eta<\eta_{\rm crit}$. The ground state of $^{12}$N is found to have  $\eta=\eta_{\rm crit}$.  The neutron-rich nuclei have very small ground-state $\eta$-values; hence, their normalization constants $\beta_{\ell=1,j}(\eta)$ are large (see Fig.~\ref{fig1}). An  odd-even staggering of $\eta$ -- due to pairing -- 
leads to an odd-even effect in $\beta_{\ell=1,j}(Z_a)$. The staggering in 
$\beta_{\ell j}$ is stronger for
nuclei with $T_z \geq 0$ than in the proton-rich systems with  $T_z < 0$.

The data for $sd$-shell nuclei shown in Fig.~\ref{fig3} exhibit similar behavior.
Most of the particle-stable $sd$-shell nuclei belong to the class $\eta<\eta_{\rm crit}$. The only system that falls decisively  into the regime of weak binding is the 5/2$^+$ ground state of $^{23}$Al, which has a small separation energy of $S_p$=141\,keV and a very large value of $\beta_{\ell,j}$ (see Ref.~\cite{Banu11} for a recent discussion). Other nuclei with 
$\eta>\eta_{\rm crit}$ are $^{29}$P, $^{32}$Cl, and $^{36,37}$K whereas  $^{17}$F, $^{30}$S, $^{33}$Cl, and $^{37}$Ca  are situated at the critical line $\eta_{\rm crit}(Z_a)$. 
A pronounced  odd-even staggering of  $\eta$ and $\beta_{\ell,j}$  is seen for both proton-rich and neutron-rich systems. The values of SPANCs  for proton-rich nuclei remain close to $\beta_{\rm crit}$.

\begin{figure}[hbt]
\center
\includegraphics[width=0.45\textwidth]{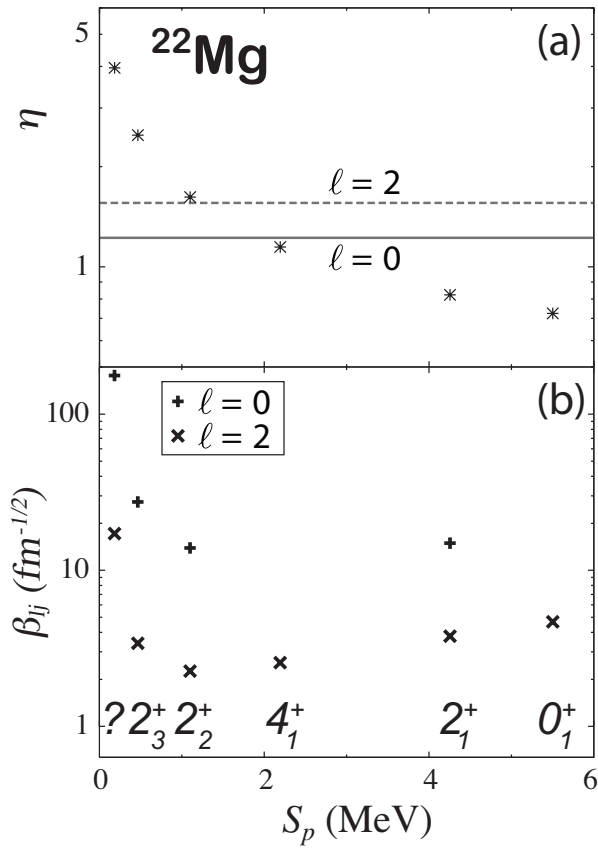}
\caption{\label{fig4} Experimental values of $\eta$ (a) and $\beta_{\ell,j}$ (b) for the low-energy states of $^{22}$Mg as a function of the proton separation energy $S_p$. The values of   $\eta_{\rm crit}$  in channels with $\ell=0$  and $\ell=2$ are indicated.}
\end{figure}
The  transition from   $\eta<\eta_{\rm crit}$ to  $\eta>\eta_{\rm crit}$ can be further explored by studying different particle-stable excited states in one nucleus, having different values of $\eta$,  populated in the capture reaction. Figure~\ref{fig4}(a)
shows experimental values of $\eta$ as a function of the proton separation energy for several excited states of   $^{22}$Mg.  A transition  from $\eta<\eta_{\rm crit}$ to $\eta>\eta_{\rm crit}$ takes place  between the  $J_i^{\pi}=4_1^+$ and $2_2^+$ levels. The corresponding SPANCs are displayed in  Fig.~\ref{fig4}(b). Those  with   $\ell=2$ show a minimum at  the $2_2^+$ state and the same is seen  in the  $\ell=0$ channel. (One may notice the  absence of $\ell=0$ data points for the  $0_1^+$ and $4_1^+$ levels as these states  cannot  be populated by the $\ell=0$ proton capture.)

\subsection{General  properties of  SPANCs for neutrons}
\label{sec:ANC-neutron}

The s.p.~neutron wave function on the asymptotic region is
\begin{equation}
u_{\ell j}(r) \simeq \beta_{\ell j} \,(-i^\ell) ~\kappa r~ h_\ell^+(i\kappa r),
\end{equation}
where 
$h_\ell^+$ is the spherical Hankel function \cite{Abr70,Gra80}.
In the limit of $\kappa \rightarrow 0$
\begin{equation}
(-i^\ell) ~\kappa r~ h_\ell^+(i\kappa r) \simeq 
\frac{(2 \ell - 1)!!}{(\kappa r)^\ell}~~{\rm for} ~\ell \ne 0.
\end{equation}
Using the same arguments as in Sec.~\ref{SPANCp1}, we conclude that 
close to the neutron threshold the neutron SPANC behaves as
\begin{equation}
\beta_{\ell j} \propto  \kappa^\ell~~{\rm for} ~\ell \ne 0.
\end{equation}
To discuss the  special case of $\ell=0$, we follow the analysis
of Ref.~\cite{Rii92}. The asymptotic part of the wave function for the $s$-wave is 
\begin{equation}\label{elzero}
u_{\ell j}(r) \simeq \beta_{\ell j} \exp(-\kappa r).
\end{equation}
Since the  external norm ${\cal N}_{\rm ext}$ (\ref{extN}) of (\ref{elzero}) is finite, the near-threshold divergence of $\int_{R_f}^{+\infty} \exp(-2\kappa r)\,dr \propto \kappa^{-1}$ must be compensated by the $\kappa$-dependence of SPANC:
\begin{equation}
\beta_{\ell=0 j=1/2} \propto  \sqrt{\kappa}.
\end{equation}

In the limit of the large binding energy, the neutron wave function exhibits the exponential behavior:
\begin{equation}\label{uneut1}
u_{\ell j}(r) \simeq \beta_{\ell j} \exp (-\kappa r).
\end{equation}
Using the reasoning of Sec.~\ref{SPANCp2}, we conclude that
$\beta_{\ell j}$ has to increase  with $\kappa$. Consequently, in the neutron case, $\beta_{\ell j}$   should monotonically decrease   with $\kappa$ 
all the way down to zero.

To describe $\beta_{\ell j}$ for neutrons in a wider energy range, one can employ the power series expansion in $\kappa$: 
\begin{subequations}\label{eq17}
\begin{IEEEeqnarray}{rcll}
\beta_{\ell=0,1/2} & = & \sum_{i=0}^{\infty}a_{\ell=0,1/2}^{(i)}\kappa^{(i+1/2)} &~~{\rm for} ~\ell = 0,\\
\beta_{\ell,j} & = & \sum_{i=0}^{\infty}a_{\ell,j}^{(i)}\kappa^{(i+\ell)} &~~{\rm for} ~\ell \ne 0.
\end{IEEEeqnarray}
\end{subequations}

\begin{figure}[hbt]
\center
\includegraphics[width=0.35\textwidth]{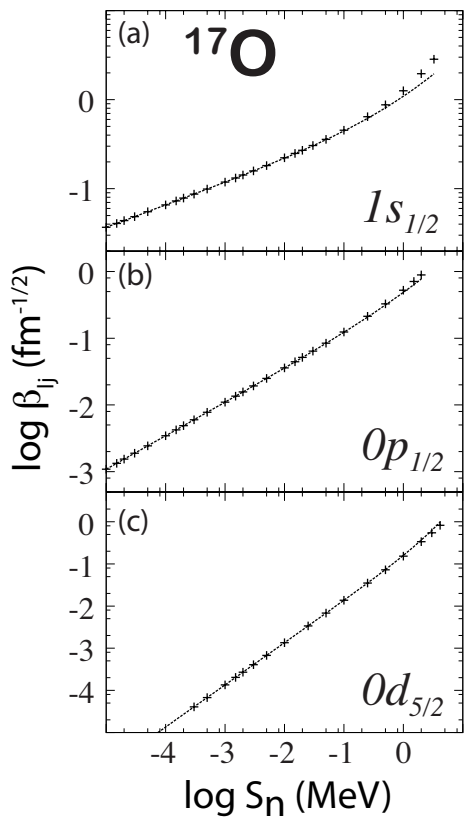}
\caption{\label{figx}$\beta_{\ell j}$ for the bound neutron states $1s_{1/2}$ (a), $0p_{1/2}$ (b), and $0d_{5/2}$ (c) in $^{17}$O as a function of  $S_n$. The fits using  expansions (\ref{eq17}) are  shown by dotted lines. The same WS potential as in Fig.~\ref{fig5} was used.}
\end{figure}

Figure~\ref{figx} shows  $\beta_{\ell j}$ for the bound $1s_{1/2}$, $0p_{1/2}$, and $0d_{5/2}$ neutron   s.p.~wave functions. The short-dashed lines are fits with two leading terms to expansions (\ref{eq17}).
In general, the coefficients $a_{\ell,j}^{(i)}$ in Eq.~(\ref{eq17}) depend on the choice of an average potential. 
The case $\ell=0$ is special. Since ${\cal N}_{\rm ext}\rightarrow 1$ in the limit $S_n\rightarrow 0$, SPANC becomes potential-independent. In particular, the coefficient $a_{\ell=0,1/2}^{(0)}$  in Eq.~(\ref{eq17}) becomes $\sqrt{2}$ in the limit of zero binding.

For $\ell=1, j=1/2$, the values of  $a_{\ell,j}^{(0)}$ are 1.549, 1.612, and 1.681 for  $d$ equal to 0.46 fm, 0.58 fm, 0.7 fm, respectively. Variations of $a_{\ell,j}^{(0)}$ with $d$ become insignificant if the r.m.s.~radius of the potential  is kept constant.

\section{Continuum Shell Model description of ANCs in mirror systems}
\label{sec:CSM}

An extensive analysis of proton and neutron ANCs for light mirror nuclei has been performed using SM and cluster model  wave function and various effective $NN$ interactions  \cite{timo98, *timo03a, timo03, *Timdes05a, Timdes05, Timdes07, Tit11}. It has been found that  the ratio of proton  and neutron  ANCs for mirror nuclei,
\begin{equation}\label{R_ANCs}
{\cal R}=|C_p/C_n|^2,
\end{equation}
is rather insensitive to model details and can be well approximated by the expression
\begin{equation}\label{Ras}
{\cal R}\approx {\cal R}_0 \equiv \left| 
\frac{e^{i \sigma_\ell(-i \eta)} F_{\ell,-i \eta}(i\kappa_p R_f)}{\kappa_p R_f j_\ell(i\kappa_n R_f)}\right|^2,
\end{equation}
where $\sigma_\ell(-i \eta)$ is the Coulomb phase shift associated with the imaginary momentum $i \kappa_p$, 
and $F_{\ell,-i \eta}$ and $j_\ell$ are the regular Coulomb and
regular Bessel functions, respectively. It is to be noted that our expression for ${\cal R}_0$ differs from that of Refs.~\cite{timo03,Timdes05} because of different definitions of complex Coulomb wave functions. In our work  we follow the convention of Refs.~\cite{Thompson86,Michel07}, which results in the appearance of an additional factor $e^{i \sigma_\ell(-i \eta)}$ in Eq.~(\ref{Ras}).

We shall verify these findings in the CSM. As  ANCs are impacted by  the configuration mixing and continuum coupling through the spectroscopic amplitudes $S_{\ell j}^{1/2}$ (\ref{eq2}),
the OQS framework of the CSM is particularly well-suited for the description of the ANCs in mirror nuclei.  Indeed, since the one-nucleon separation energies in mirror systems can be appreciably different, the particle continuum, both of a resonant and non-resonant character, can  impact   properties of states involved, especially when dealing with near-threshold energies.

The principal difficulty in the formulation of the CSM is the treatment of the unbound space of states, i.e., resonances  and the non-resonant continuum \cite{oko, micrev, Mic09}. Therefore, whatever strategy  is adopted to formulate the configuration interaction approach for OQSs, the key points are: (i) the treatment of the  s.p.~continuum, and (ii) the definition of the many-body Fock space.

Historically, the first approach to formulate the CSM in Hilbert space was based on the projection technique \cite{fesh1, *fesh2}. Here, the Hilbert space is divided into orthogonal subspaces ${\cal Q}_{i_{\rm cont}}$ containing different numbers ${i_{\rm cont}}$ (${i_{\rm cont}}=0,1,\cdots$) of particles  in the scattering continuum. An OQS description of ${\cal Q}_0$  space includes couplings to the environment of decay channels through the energy-dependent effective Hamiltonian \cite{Bar77, *Rot78, Ben99, *Ben00, *Rot05, *Rot06a, Vol05, *Vol06}:
\begin{equation}
{\cal H}_{{\cal Q}_0{\cal Q}_0}(E)=H_{{\cal Q}_0{\cal Q}_0}+W_{{\cal Q}_0{\cal Q}_0}(E),
\label{eq21}
\end{equation}
where $H_{{\cal Q}_0{\cal Q}_0}$ is the standard SM Hamiltonian describing the internal dynamics in the closed quantum system approximation and $W_{{\cal Q}_0{\cal Q}_0}(E)$ is the energy-dependent continuum coupling term. 
The Hamiltonian (\ref{eq21}) is Hermitian below the first particle emission threshold and complex-symmetric above it. 

The s.p.~resonances have to be regularized before including them in the subspace of discrete states. An effective way of doing this is based on the anamneses of resonances, i.e., images of resonances in the space of ${\cal L}^2-$functions  \cite{faes}. Bound s.p.~states, resonance anamneses, and real-energy scattering continuum states form a complete s.p.~basis \cite{faes}. The many-body Hilbert space can be represented by Slater determinants spanned by this basis. In practical applications of the real-energy CSMs, such as  SMEC, one uses  phenomenological arguments to restrict the number of particles in the scattering continuum. Technical difficulties associated with the correct treatment of the multiparticle channel wave functions rapidly grow with $i_{\rm cont}$; hence,
in practical applications, the  number of particles in the scattering continuum  has so far not exceeded two \cite{Ben99, *Ben00, *Rot05, *Rot06a}. 

Recently, a different strategy based on the rigged Hilbert space formulation of quantum mechanics \cite{Gel61, *Boh78, *Lud83} has resulted in the complex-energy CSM (the GSM) \cite{Mic02, *Mic03, *IdB02, *Rot06, *Rot09, PRC_O_He_2003, PRC_Lithium_2004}, which is a natural generalization of the standard SM for unbound systems. In this formulation, the maximum number of particles in the scattering continuum is not  {\it a priori} prescribed, but follows from the Schr\"{o}dinger variational principle for the many-body Hamiltonian. The  s.p.~GSM  basis is given by the Berggren ensemble \cite{berggren, *Ber93, *Lin93}, which consists of Gamow (resonant) states and the non-resonant continuum. (For a detailed description of the GSM, see Ref. \cite{micrev}.)  The GSM Hamiltonian is Hermitian. However, since the s.p.~vectors  have either outgoing or scattering  asymptotics,  the GSM Hamiltonian matrix is complex symmetric and its eigenvalues are complex above the particle emission threshold. Hence, both real-energy and complex-energy  CSM formulations lead to a non-Hermitian eigenvalue problem above the  threshold. 

\subsection{Theoretical framework}\label{fram}

In both GSM and SMEC, we assume  that the nucleus can be
described as a system of $A_{val}=Z_{val}+N_{val}$ valence
nucleons moving around a closed  core $A_c=Z_c+N_c$.

\subsubsection{GSM  framework}\label{fram-GSM}

The translationally invariant
GSM Hamiltonian, written in 
intrinsic nucleon-core
coordinates of the cluster-orbital shell model \cite{Ikeda},
can be written as:
\begin{equation}
H= \sum_{i=1}^{A_{val}}\left [ \frac{p_{i}^{2}}{2\mu_i} + U_{i} \right] + \sum_{i<j}^{A_{val}} \left[ V_{ij} 
+ \frac{1}{M_c} \gras{p}_{i}\gras{p}_{j} 
\right],
\label{GSM_Hamiltonian}
\end{equation}
where $\mu_i$ is the reduced mass of either the proton or neutron ($1 / \mu_i =1/m_i + 1/M_{c}$),
$U_{i}$ is  the s.p.~potential describing the field of the
core, $V_{ij}$ is the two-body residual interaction between
valence nucleons,
and the last term represents the two-body energy recoil. 
The particle-core interaction is a sum of nuclear and Coulomb terms:
\begin{equation}
U_i = U_i^N + U_i^C.
\label{GSM_Hsp}
\end{equation}
The nuclear potential $U_i^N$ is approximated by a Woods-Saxon (WS) field with a spin-orbit term \cite{Mic03}, and the Coulomb field $U_i^C$ is
generated by a Gaussian density of $Z_c$ core protons \cite{Mic10}. Similarly,
the residual interaction can split into nuclear and Coulomb parts:
\begin{equation}
V_{ij} = V_{ij}^N + V_{ij}^C,
\label{GSM_V}
\end{equation}
where $V^N$  is the modified surface Gaussian interaction (MSGI) \cite{Mic10} and $V^C$ is the two-body  Coulomb interaction that requires special treatment due to its infinite range \cite{Mic10,PRC_Coulomb_2011}.
Namely, $V^C$ is rewritten as:
\begin{equation}\label{UCoul}
U^C_{Z_{val}-1} + \left[V^C -  U^C_{Z_{val}-1}\right]^{HO},  
\end{equation}
where $U^C_{Z_{val}-1}$ takes care of the long-range asymptotic behavior of the Coulomb interaction. As $U^C_{Z_c} + U^C_{Z_{val}-1} = U^C_{Z-1}$, the long-range physics of the Coulomb Hamiltonian is treated almost exactly.
The second term in Eq.~(\ref{UCoul}) and the two-body  recoil term
can be expanded in a harmonic oscillator (HO) basis \cite{PRC_real_inter_2006,Mic10}.  In our calculations, we took nine HO shells with the oscillator length 
$b = 2$\,fm.

The radial overlap integrals were calculated using Eq.~(\ref{eqex}), where
the sum over ${\cal B}$ states runs over the complete Berggren ensemble; hence, the result is independent of the s.p.~basis representation. The ANC is  obtained directly from  Eq.~(\ref{eq1}). 

The GSM calculations presented in this paper were carried out for 
$p$-shell systems
$^6$Li/$^7$Be ($\equiv$$^6$Li+$p$ $\rightarrow$ $^7$Be), $^6$Li/$^7$Li
($\equiv$$^6$Li+$n$ $\rightarrow$ $^7$Li), and $^7$Be/$^8$B, $^7$Li/$^8$Li (assuming  $^4$He core), and $sd$-shell nuclei
$^{16}$O/$^{17}$F and $^{16}$O/$^{17}$O (assuming $^{12}$C core). 

As the protons (neutrons) are well bound in $^7$Be and $^8$Li ($^7$Li and $^8$B), only  proton (neutron)  bound s.p.~shells are included in the model space, namely the $0p_{3/2}$ and $0p_{1/2}$ states. On the other hand, the full $p_{3/2}$ and $p_{1/2}$ Berggren basis, consisting of resonant and scattering states,  is taken into account for the neutron (proton) space. In this way, completeness is assured in the neutron (proton) $p$-space. This is especially important for the ground state of the proton halo nucleus  $^8$B.

Since the precise description of reaction thresholds is crucial for precise determination  of   radial overlap integrals,
the parameters defining the WS potential and MSGI interaction
have been fitted independently for each pair of
nuclei considered to reproduce the separation energy and lowest excited states of the heavier (parent)  nucleus. The  WS potential diffuseness  has been assumed to be  $d=0.65$\,fm in all cases; WS radius has been set to $R_0=2$\,fm for $A=6-8$ nuclei
and $R_0=2.8$\,fm for $A=16/17$ pairs; and spin-orbit strength was taken as 7.5 MeV for $A=6-8$ nuclei,
and 7.920 MeV (proton) and 8.463 MeV (neutron) for $A=16/17$ pairs. 
For  $A=6/7$ pairs, the WS strength  is 45.455 MeV for $^6$Li/$^7$Be (protons) and  $^6$Li/$^7$Li (neutrons),
and  55 MeV for $^6$Li/$^7$Li (protons) and  $^6$Li/$^7$Be (neutrons).
For $A=7/8$ pairs, the corresponding values are 46.273 MeV (protons) and 65 MeV (neutrons) for $^7$Be/$^8$B,
and 75 MeV (protons) and 45.799 MeV (neutrons) for $^7$Li/$^8$Li.
For  $A=16/17$ pairs, the WS strength is  46.427 MeV  and 46.034 MeV for protons and neutrons, respectively.
The coupling strengths of MSGI depend on the quantum numbers 
$J$ and $T$ of the nucleon pair 
\cite{Mic10}. The $T=0$ constants $V^N_{J,T=0}$  are  listed in
Table~\ref{MSGI_parameters} and the   $T=1$ constants are given by 
$V^N_{J,T=1}=0.5V^N_{J,T=0}$.
\begin{table}[htb]
\begin{center}
\caption{\label{MSGI_parameters}Coupling constants $V^N_{J,T=0}$ of the MSGI residual interaction (in MeV fm$^3$) \cite{Mic10} for the considered  pairs of nuclei  and all allowed values of $J$.
The $J$=4 and 5 coupling constants in $A=16/17$ systems have been set to $-9.0$.
}
\begin{ruledtabular}
\begin{tabular}{c|cccc}
& $J=0$ &  $J=1$ & $J=2$ & $J=3$  \\ \hline  \\ [-1.5ex]
$^6$Li/$^7$Be      & $-$34.000 & $-$23.200 & $-$24.300 & $-$17.384  \\
$^6$Li/$^7$Li      & $-$34.000 & $-$23.200 & $-$24.000 & $-$17.000  \\
$^7$Be/$^8$B       & $-$25.593 & $-$16.007 & $-$10.288 & $-$12.400   \\
$^7$Li/$^8$Li      & $-$24.662 & $-$14.957 & $-$10.592 & $-$12.400   \\
$^{16}$O/$^{17}$F  & $-$16.540 & $-$12.564 & $-$5.870  & $-$9.000   \\
$^{16}$O/$^{17}$O  & $-$16.540 & $-$12.564 & $-$5.870  & $-$9.000   
\end{tabular}
\end{ruledtabular}
\end{center}
\end{table}

For $A=16/17$ systems, the model space is the same for protons and neutrons: it consists of the resonant shells $0p_{1/2}$, $0d_{5/2}$, and $1s_{1/2}$, and
the  $s_{1/2}$ and $d_{5/2}$ scattering continua. The $p_{1/2}$ and $d_{3/2}$ 
partial waves are discarded as they do not impact the asymptotic behavior of   the $0_1^+$ ground state of $^{16}$O and the $5/2_1^+$ and $1/2_1^+$  states of $^{17}$O and $^{17}$F. 
In the pole approximation, in which only resonant states are considered,  amplitudes of 2p-2h and 4p-4h $0p_{1/2} \rightarrow sd$ excitations are  of the order of $10^{-3}$ and  $10^{-6}$, respectively. Consequently,
as configuration mixing effects in the states considered are  weak,  only  two particle-two hole excitations have been  allowed from the proton and neutron $0p_{1/2}$ states, and only one particle has been allowed to occupy $s_{1/2}$ and $d_{5/2}$ scattering states. 
In all cases, scattering contours have been discretized utilizing a Gauss-Legendre quadrature. 
We have checked that a 45-point discretization for $A=6/7$ (and 30-point discretization for other systems)  is sufficient to maintain  a correct asymptotic behavior of $I^a_{bc;\ell j}(r)$  up to at least $\sim 8$ fm.
For the s.p.~basis, we took a  Gamow Hartree-Fock (GHF) ensemble \cite{PRC_Lithium_2004} corresponding to a parent nucleus. Sphericity of the GHF potential is guaranteed by the use of the uniform filling approximation.
By taking the GHF basis  we minimize configuration mixing. The  many-body GSM states have been determined by a diagonalization of the GSM matrix using the Davidson method extended to complex-valued Hamiltonians. The identification of the outgoing GSM states has been carried out by applying the overlap method \cite{PRC_O_He_2003}.

\begin{table}[htb]
\begin{center}
\caption{\label{fitted_separation_energies}Calculated and experimental
separation energies for the considered pairs of nuclei.
}
\begin{ruledtabular}
\begin{tabular}{c|cccc}
& $J^{\pi}(A-1)$ & $J^{\pi}(A)$ & $E_{\rm GSM}$ (MeV) & $E_{\rm exp}$ (MeV)\\ \hline  \\ [-1.5ex]
$^6$Li/$^7$Be     & $2^+$ &  $3/2^-$ &  5.606    &  5.606    \\
$^6$Li/$^7$Li     & $2^+$ &  $3/2^-$ &  7.250    & 7.250    \\
$^7$Be/$^8$B      & $3/2^-$ &  $2^+$ &  0.137    &  0.137    \\
$^7$Li/$^8$Li     & $3/2^-$ &  $2^+$  &  2.034    &   2.033   \\
$^{16}$O/$^{17}$F & $0^+$ &  $5/2^+$  &  0.601   & 0.600  \\
$^{16}$O/$^{17}$O & $0^+$ &  $5/2^+$  &  4.143    &  4.143   
\end{tabular}
\end{ruledtabular}
\end{center}
\end{table}
The parameter optimization was carried out  using the multidimensional Newton method. The fine-tuning was done manually to  adjust the thresholds to experimental value with a 1 keV precision. For the excited states, we reproduce experimental data  with a precision of a few tens of keV. 
The calculated  separation energies are compared to experiment in Table \ref{fitted_separation_energies}, while Table~\ref{fitted_energies_widths}
displays 
excitation energies and widths of the lowest excited states in the nuclei considered.
\begin{table}[htb]
\begin{center}
\caption{\label{fitted_energies_widths}Calculated and experimental excitation energies (in MeV) and widths (in keV) of the first excited states of the parent nucleus. The approximate width $\Gamma^{\rm (app)}_{\rm GSM}$ of  Eq.~(\ref{narrow_width_formula}) is shown for comparison. 
}
\begin{ruledtabular}
\begin{tabular}{c|cccccc}
& $J^{\pi}$ &  $E_{\rm GSM}$ & $E_{\rm exp}$ & $\Gamma_{\rm GSM}$ & $\Gamma^{\rm (app)}_{GSM}$ & $\Gamma_{\rm exp}$ \\ \hline  \\ [-1.5ex]
$^7$Be     & $1/2^-$ &  0.439    & 0.429  & 0 & 0 & 0   \\
$^7$Li     & $1/2^-$ &  0.459    & 0.478  & 0 & 0 & 0    \\
$^8$B      & $1^+$ &  0.771      & 0.770  & 24 & 23 & 36     \\
$^8$B      & $3^+$ &  2.278      & 2.320  & 275 & 258 & 350   \\
$^8$Li     & $1^+$  &  0.992     & 0.981  & 0 & 0 & 0    \\
$^8$Li     & $3^+$  &  2.222     & 2.255  & 21 & 21 & 32    \\
$^{17}$F   & $1/2^+$  &  0.495   & 0.495  & 0 & 0 & 0   \\
$^{17}$O   & $1/2^+$  &  0.870   & 0.871  & 0 & 0 & 0    
\end{tabular}
\end{ruledtabular}
\end{center}
\end{table}

\subsubsection{SMEC  framework}\label{fram-SMEC}

In this study, the scattering environment is provided by one-nucleon decay channels, i.e.\ we solve the Schr\"{o}dinger equation in the function space ${\cal Q}_0\oplus{\cal Q}_1$.
The SMEC solutions in ${\cal Q}_0$ are found by solving the eigenproblem for the non-Hermitian Hamiltonian (\ref{eq21}):
\begin{eqnarray}
{\cal H}_{{\cal Q}_0{\cal Q}_0}|\Psi_{\alpha}\rangle&=&{\cal E}_{\alpha}(E,V_0)|\Psi_{\alpha}\rangle \nonumber \\
\langle \Psi_{\bar \alpha}|{\cal H}_{{\cal Q}_0{\cal Q}_0}&=&{\cal E}_{\alpha}^*(E,V_0)\langle \Psi_{\bar \alpha}|
\label{eqop2}
\end{eqnarray}
in the biorthogonal basis: $\langle \Psi_{\bar \alpha}|\Psi_{\beta}\rangle=\delta_{\alpha\beta}$. As usual, left $|\Psi_{\alpha}\rangle$ and right $|\Psi_{\bar \alpha}\rangle$ eigenvectors  are related by the complex conjugation. 
In Eq.~(\ref{eqop2}), $E$ and $V_0$  stand for a scattering energy and a (real) continuum coupling constant in the coupling terms ${\cal H}_{{\cal Q}_0,{\cal Q}_1}$ and ${\cal H}_{{\cal Q}_1{\cal Q}_0}$:
\begin{equation}
W_{{\cal Q}_0{\cal Q}_0}(E)=H_{{\cal Q}_0{\cal Q}_1}G_{{\cal Q}_1}^{(+)}(E)H_{{\cal Q}_1{\cal Q}_0},
\label{eqop4}
\end{equation}
where $G_{{\cal Q}_1}^{(+)}(E)$ is the one-nucleon Green's function. The energy scale is defined by the position of the one-nucleon emission threshold. At resonance, the eigenvalue of the effective Hamiltonian can be identified with the narrow pole of the scattering matrix (the $S$-matrix). 

Inside of the interaction region, dominant contributions to the full solution of the Schr\"{o}dinger equation in ${\cal Q}_0\oplus{\cal Q}_1$ are given by the eigenfunctions of ${\cal H}_{{\cal Q}_0{\cal Q}_0}(E)$. This is the main reason why eigenfunctions of the non-Hermitian effective Hamiltonian are essential to understand properties of the OQS. The SMEC eigenvectors
$\Psi_{\alpha}$  are  related to the eigenstates $\Phi_j$ of the closed quantum system Hamiltonian $H_{{\cal Q}_0{\cal Q}_0}$ by a linear orthogonal transformation: 
\begin{equation}
\Psi_{\alpha}=\sum_jb_{\alpha j}\Phi_j.
\label{eqop3}
\end{equation}

In our SMEC calculations, for the  effective SM Hamiltonian  $H_{{\cal Q}_0{\cal Q}_0}$  we took  the Cohen-Kurath interaction \cite{Cohen}  for $A=6-8$ systems and the ZBM effective interaction \cite{ZBM}  for $sd$-shell nuclei. The continuum-coupling term (\ref{eqop4}) was approximated by means of
the Wigner-Bartlett contact interaction:
\begin{equation}
V_{12}=V_0\left[\alpha+(1-\alpha)P_{12}^{\sigma}\right]\delta\left({\bf r_1}-{\bf r_2}\right),
\label{WB}
\end{equation}
where $P_{12}^{\sigma}$ is the spin exchange operator and $\alpha=0.73$.
The magnitude of the continuum coupling varies depending on the structure of SM wave function in a target nucleus with 
$(A-1)$ nucleons and the energy of the lowest one-nucleon emission threshold, which is fixed at the experimental value in all calculations.
SMEC is particularly suited for studies of the qualitative effects of the continuum coupling because the strength of this coupling can be changed continuously from a SM limit ($V_0=0$) to  physically relevant values.

The expectation value of any operator ${\hat O}$ can be calculated as:
\begin{eqnarray}
\langle{\hat O}\rangle=\langle{\Psi}_{{\bar \alpha}}|{\hat O}|{\Psi}_{\alpha}\rangle ~ \ ,
\label{eqop5}
\end{eqnarray}
In case of the spectroscopic factor one has:
\begin{eqnarray}
{\hat O}=a^{\dagger}|t\rangle\langle t|a,
\end{eqnarray}
where $|t\rangle$ is the target state of the $(A-1)$-system. 
For a single SM configuration, the ANC is proportional to the square root of the spectroscopic factor (\ref{eq4}).
In SMEC, the spectroscopic factors depend on the total energy $E$ of the system and exhibit characteristic near-threshold variations that depend on the transferred angular momentum. In the multichannel representation of a many-body system, the flux conservation  imposes an intricate interdependence between various spectroscopic factors not only on $E$ but also on the strength of the continuum coupling $V_0$. This salient dependence of the ANC on the continuum coupling strength is a quantal effect, beyond the generic features discussed in Sec.~\ref{sec:ANC}. 

In the  case of multi-channel coupling, the squared  norm (\ref{Cnorm}) becomes
\begin{eqnarray}
\left|C\right|^2=\sum_{\ell,j} \left|\beta_{\ell j}\right|^2 S_{\ell j}.
\label{a4}
\end{eqnarray}
The ratio of proton  and neutron  ANCs for mirror nuclei (\ref{R_ANCs}) can be directly computed by means of Eq.~(\ref{a4})  applied to
charged-particle and neutral-particle
radiative capture reactions.

\subsection{CSM description of ANCs for bound states}

In this section, we discuss the ANCs corresponding to single-nucleon 
capture reactions  between bound states.
We first present our GSM results ($E_{CM}=0$).  The ANCs in GSM can be directly extracted from the calculated radial overlap integrals  by fitting their tail to Whittaker functions at large values of $r$ (=7-8\,fm).
We can use such a direct method of extraction because the asymptotic behavior of $I^a_{bc;\ell j}(r)$  is well controlled in GSM. 

\begin{figure}[hbt]
\center
\includegraphics[width=0.35\textwidth]{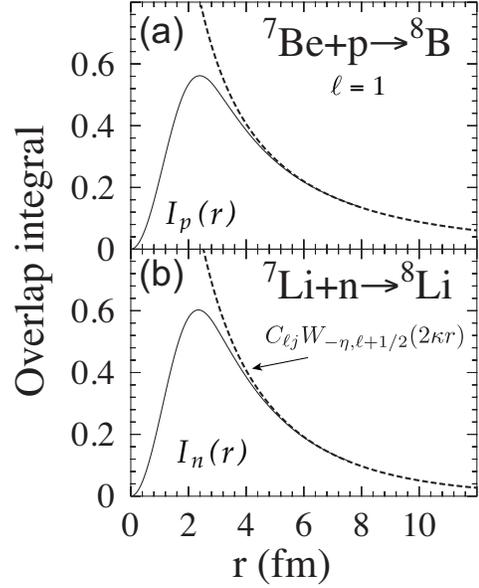}
\caption{\label{Ov1-GSM}Radial overlap integrals 
$I^a_{bc;\ell j}(r)$ calculated in GSM for $^7{\rm Be}_{3/2^-}+p\rightarrow {^8{\rm B}}_{2^+}$ (a) and $^7{\rm Li}_{3/2^-}+n\rightarrow {^8{\rm Li}}_{2^+}$ (b) for $\ell=1$ and $j=3/2$. The tail of the radial overlap integral is fitted by the Whittaker function (dashed line) to extract ANC.}
\end{figure}
Figure~\ref{Ov1-GSM}  shows how this procedure works for the  radial overlap integrals corresponding to  $\ell=1$ ($0p_{1/2}$ and $0p_{3/2}$) protons and neutrons
in the $2^+$ ground states of the mirror nuclei $^8$B, $^8$Li. This example is non trivial as the  configuration mixing  is appreciable, and the g.s.~of $^8B$ is a proton halo -- as seen from the extended tail of the overlap function
in Fig.~\ref{Ov1-GSM}(a). 

\begin{figure}[hbt]
\center
\includegraphics[width=0.35\textwidth]{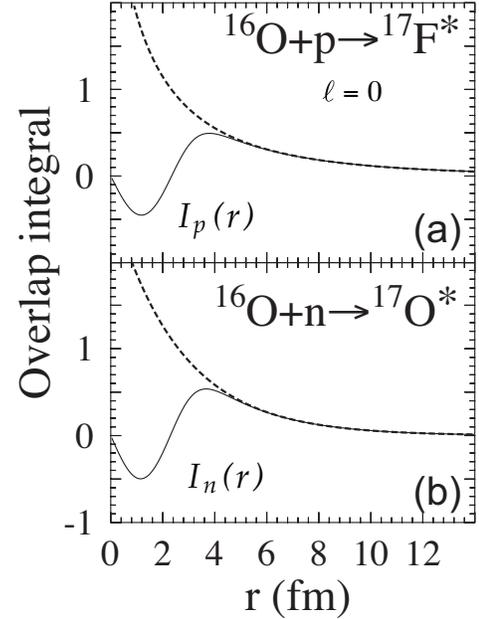}
\caption{\label{Ov2-GSM}Similar as in Fig.~\ref{Ov1-GSM} but for $^{16}{\rm O}_{0^+}+p\rightarrow {^{17}{\rm F}}^*_{1/2^+}$ (a) and  $^{16}{\rm O}_{0^+}+n\rightarrow {^{17}{\rm O}}^*_{1/2^+}$ (b) for $\ell=0$.}
\end{figure}
The second example presented in Fig.~\ref{Ov2-GSM}
pertains to the first excited  $J^{\pi}=1/2_1^+$ subthreshold halo state  in $^{17}$F and $^{17}$O, as seen in  Table~\ref{fitted_energies_widths}. 
Here, both proton and neutron overlap functions are very extended. Nevertheless, the extraction of the ANCs does not cause any problems.

\begin{table}[htb]
\begin{center}
\caption{\label{tablo_fe}GSM predictions for ANCs in mirror systems    compared to VMC results \cite{Nol11} and  experimental data \cite{Goncharov,Trache,Tabacaru,Burzynski,timo03}.  For 
$^6{\rm Li}_{1^+}+n\rightarrow {^7{\rm Li}}_{3/2^-}$,  two experimental values have been reported \cite{Goncharov}. 
For  $A=6-8$ systems, we show individual  contributions from $p_{1/2}$ and $p_{3/2}$ channels and their Hermitian norm (\ref{Cnorm}).
}
\begin{ruledtabular}
\begin{tabular}{c|cccc}
Overlap & $j^{\pi}$ &  GSM  & VMC  & Exp. \\ \hline  \\ [-1.5ex]
\multirow{3}{*}{$^6{\rm Li}_{1^+} \xrightarrow{p} {^7{\rm Be}}_{3/2^-}$} 
                    & 1/2$^-$ & 0.431  & 1.870  & --- \\
                    & 3/2$^-$ & 1.499  & 2.150  & --- \\
            & 1/2$^-$+3/2$^-$ & 1.559  & 2.850  & --- \\[5pt]  
\multirow{4}{*}{$^6{\rm Li}_{1^+} \xrightarrow{n} {^7{\rm Li}}_{3/2^-}$} 
                    & 1/2$^-$ & 0.422 & 1.652  & --- \\
                    & 3/2$^-$ & 1.456 & 1.890  & --- \\
           & \multirow{2}{*}{1/2$^-$+3/2$^-$} & \multirow{2}{*}{1.559}  & \multirow{2}{*}{2.510}  & 1.86 $\pm$ 0.06 \\
           &                 &        &        & 2.57 $\pm$ 0.06 \\[5pt]   
\multirow{3}{*}{$^7{\rm Be}_{3/2^-} \xrightarrow{p} {^8{\rm B}}_{2^+}$} 
                    & 1/2$^-$ & 0.049 & 0.246  &  0.23 $\pm$ 0.01 \\
                    & 3/2$^-$ & 0.765 & 0.691  & 0.64 $\pm$ 0.03  \\
           & 1/2$^-$+3/2$^-$ & 0.767  & 0.733  & 0.68 $\pm$ 0.04 \\ [5pt]                   
\multirow{3}{*}{$^7{\rm Li}_{3/2^-} \xrightarrow{n} {^8{\rm Li}}_{2^+}$} 
                    & 1/2$^-$ & 0.041 & 0.218  &  0.22 $\pm$ 0.01 \\
                    & 3/2$^-$ & 0.750 & 0.618  & 0.62 $\pm$ 0.03  \\
           & 1/2$^-$+3/2$^-$ & 0.752  & 0.655  & 0.66 $\pm$ 0.03 \\ [5pt]   
\multirow{2}{*}{$^{16}{\rm O}_{0^+} \xrightarrow{p} {^{17}{\rm F}}_{j^\pi}$} 
                    & 5/2$^+$ & 0.880  & ---  & 0.95 $\pm$ 0.09 \\
                    & 1/2$^+$ & 73.74  & ---  & --- \\  [5pt]   
\multirow{2}{*}{$^{16}{\rm O}_{0^+} \xrightarrow{n} {^{17}{\rm O}}_{j^\pi}$} 
                    & 5/2$^+$ & 0.805  & ---  & 0.82 $\pm$ 0.01 \\
                    & 1/2$^+$ & 2.785  & ---  & ---  
\end{tabular}
\end{ruledtabular}
\end{center}
\end{table}
The GSM predictions for ANC are listed in Table~\ref{tablo_fe} together with the  Variational Monte Carlo (VMC) results of Ref.~\cite{Nol11} 
and experimental data.
One can see  that the values of ANC for the $p_{1/2}$ proton and neutron partial waves  are very different in GSM and VMC.
While  VMC values of ANCs in $p$-shell nuclei are usually closer to  experiment for individual partial waves,
both GSM and VMC perform well when the norm (\ref{Cnorm})  is considered. 
The GSM results for the $5/2^+$ partial wave in the  $A=16\rightarrow 17$ capture are very close to the data. This does not come as a surprise as the associated spectroscopic factors are almost equal to one, i.e.,  the  ANC and SPANC values are practically identical.

\begin{table}[htb]
\begin{center}
\caption{\label{tablo_R}Ratio $\mathcal{R}$ (\ref{R_ANCs}) calculated in GSM and VMC \cite{Nol11}, compared with 
 estimate $\mathcal{R}_0$ (\ref{Ras}) and experiment  for the mirror systems of  Table~\ref{tablo_fe}.
}
\begin{ruledtabular}
\begin{tabular}{c|ccccc}
Mirror pair & $j^{\pi}$ &  ${\cal R}_{\rm GSM}$  & ${\cal R}_{\rm VMC}$  &  ${\cal R}_0$ &  ${\cal R}_{\rm Exp.}$ \\ \hline  \\ [-1.5ex]
\multirow{3}{*}{$({^7{\rm Be}}/{^7{\rm Li}})_{3/2^-}$} 
                    & 1/2$^-$ & 1.04  & 1.28  & 1.06 &  --- \\
                    & 3/2$^-$ & 1.06  & 1.29  & 1.06 &  ---  \\
            & 1/2$^-$+3/2$^-$ & 1.06  & 1.29  & 1.06 &  ---  \\[5pt]  
\multirow{3}{*}{$({^8{\rm Be}}/{^8{\rm Li}})_{2^+}$} 
                    & 1/2$^-$ & 1.39  & 1.27  & 1.12 &  1.08 $\pm$ 0.18 \\
                    & 3/2$^-$ & 1.04  & 1.25  & 1.12 & 1.08 $\pm$ 0.15  \\
            & 1/2$^-$+3/2$^-$ & 1.04  & 1.25  & 1.12 &  1.08 $\pm$ 0.15  \\[5pt] 
\multirow{2}{*}{$(^{17}{\rm F}/{^{17}{\rm O}})_{j^\pi}$} 
                    & 5/2$^+$ & 1.20  & ---  & 1.22 &  1.33 $\pm$ 0.20 \\
                    & 1/2$^+$ & 701  & ---  & 796 & ---            
\end{tabular}
\end{ruledtabular}
\end{center}
\end{table}
The   ratios $\mathcal{R}$ (\ref{R_ANCs}) predicted in GSM and VMC are displayed in Table~\ref{tablo_R}. They are  compared to experimental data and 
the approximate  expression  $\mathcal{R}_0$ of Eq.~(\ref{Ras}). (The predictions of microscopic cluster models can be found in Refs. \cite{Timdes05a} and \cite{Tit11}.)
As discussed in Ref.~\cite{timo03}, as compared to ANCs, values of $\mathcal{R}$  exhibit less model dependence. Overall, 
predicted ratios $\mathcal{R}$ are fairly close to  the  estimate $\mathcal{R}_0$  and experiment.

In the following, we shall discuss the mirror ANCs  using SMEC. The advantage of the projection technique used in SMEC is that the continuum coupling  can be switched off and, therefore, the effect of the environment of decay channels on the mirror ANCs can be studied separately from the effect of internal dynamics governed by the effective interaction. 
The SMEC calculations presented in this paper were
carried out for $p$- and $sd$-shell bound-state mirror reactions:  $^6$Li/$^7$Be and $^6$Li/$^7$Li;
$^7$Be/$^8$B and $^7$Li/$^8$Li;
$^{11}$C/$^{12}$N and $^{11}$B/$^{12}$B;
$^{16}$O/$^{17}$F and $^{16}$O/$^{17}$O; and
$^{17}$F/$^{18}$Ne and $^{17}$O/$^{18}$O.

\begin{table}[htb]
\begin{center}
\caption{\label{tablo_SMEC1}SMEC predictions for ANCs in mirror systems  
calculated with Cohen-Kurath ($A=7,8,12$) and ZBM ($A=17,18$) interactions for three values  of the continuum coupling strength  $V_0$ (in GeV\,fm$^3$):
0 (SM limit), $-0.65$,  and $-1.30$,  all in the physically relevant range of continuum-coupling.
}
\begin{ruledtabular}
\begin{tabular}{c|cccc}
Overlap & $j^{\pi}$ &  0 (SM)  & $ 0.65$  & $1.30$  \\ \hline  \\ [-1.5ex]
\multirow{3}{*}{$^6{\rm Li}_{1^+} \xrightarrow{p} {^7{\rm Be}}_{3/2^-}$} 
                    & 1/2$^-$ & 1.017  & 1.024  & 1.046 \\
                    & 3/2$^-$ & 1.279  & 1.282  & 1.291 \\
            & 1/2$^-$+3/2$^-$ & 1.663  & 1.641  & 1.661 \\[5pt]  
\multirow{3}{*}{$^6{\rm Li}_{1^+} \xrightarrow{n} {^7{\rm Li}}_{3/2^-}$} 
                    & 1/2$^-$ & 0.988 & 0.995  & 1.016 \\
                    & 3/2$^-$ & 1.243 & 1.246  & 1.254 \\
            & 1/2$^-$+3/2$^-$ & 1.588  & 1.595  & 1.614 \\[5pt]   
\multirow{3}{*}{$^7{\rm Be}_{3/2^-} \xrightarrow{p} {^8{\rm B}}_{2^+}$} 
                    & 1/2$^-$ & 0.164 & 0.149  &  0.088 \\
                    & 3/2$^-$ & 0.702 & 0.702  & 0.700  \\
           & 1/2$^-$+3/2$^-$ & 0.721  & 0.717  & 0.705\\ [5pt] 
\multirow{3}{*}{$^7{\rm Li}_{3/2^-} \xrightarrow{n} {^8{\rm Li}}_{2^+}$} 
                    & 1/2$^-$ & 0.160 & 0.151  &  0.115 \\
                    & 3/2$^-$ & 0.685 & 0.685  & 0.684 \\
           & 1/2$^-$+3/2$^-$ & 0.704  & 0.702  & 0.704 \\ [5pt] 
\multirow{3}{*}{$^{11}{\rm C}_{3/2^-} \xrightarrow{p} {^{12}{\rm N}}_{1^+}$} 
                    & 1/2$^-$ & 1.122 & 1.126  &  1.127 \\
                    & 3/2$^-$ & 0.535 & 0.530  & 0.527  \\
           & 1/2$^-$+3/2$^-$ & 1.242  & 1.244 & 1.244 \\ [5pt]            
\multirow{3}{*}{$^{11}{\rm B}_{3/2^-} \xrightarrow{n} {^{12}{\rm B}}_{1^+}$} 
                    & 1/2$^-$ & 0.971 & 0.975  &  0.977 \\
                    & 3/2$^-$ & 0.465 & 0.459  & 0.454  \\
           & 1/2$^-$+3/2$^-$ & 1.077  & 1.078 & 1.077 \\ [5pt]                  
\multirow{2}{*}{$^{16}{\rm O}_{0^+} \xrightarrow{p} {^{17}{\rm F}}_{j^\pi}$} 
                    & 5/2$^+$ & 0.869  & 0.867 & 0.863 \\
                    & 1/2$^+$ & 75.91  & 76.01  & 76.09 \\  [5pt]   
\multirow{2}{*}{$^{16}{\rm O}_{0^+} \xrightarrow{n} {^{17}{\rm O}}_{j^\pi}$} 
                    & 5/2$^+$ & 0.788  & 0.787  & 0.783 \\
                    & 1/2$^+$ & 2.771  & 2.773  & 2.776 \\  [5pt]
\multirow{3}{*}{$^{17}{\rm F}_{5/2^+} \xrightarrow{p} {^{18}{\rm Ne}}_{2^+}$} 
                    & 1/2$^+$ & 4.500  & 4.610  & 4.961 \\
                    & 5/2$^+$ & 1.739  & 1.758  & 1.702 \\                    
            & 1/2$^+$+5/2$^+$ & 4.823  & 4.934 & 5.245 \\ [5pt]      
\multirow{3}{*}{$^{17}{\rm O}_{5/2^+} \xrightarrow{n} {^{18}{\rm O}}_{2^+}$}            
                    & 1/2$^+$ & 2.801  & 2.796  & 2.794 \\                  
                    & 5/2$^+$ & 1.504  & 1.544  & 1.580 \\                      
            & 1/2$^+$+5/2$^+$ & 3.174 & 3.194 & 3.210 \\ [5pt]                                        
\multirow{3}{*}{$^{17}{\rm F}_{5/2^+} \xrightarrow{p} {^{18}{\rm Ne}}_{2^+_2}$}            
                    & 1/2$^+$ & 13.36  & 13.92 & 12.36 \\  
                    & 5/2$^+$ & 1.644  & 1.820  & 2.299 \\                      
            & 1/2$^+$+5/2$^+$ & 13.46 & 14.04 & 12.57\\ [5pt]                  
 \multirow{3}{*}{$^{17}{\rm O}_{5/2^+} \xrightarrow{p} {^{18}{\rm O}}_{2^+_2}$}            
                    & 1/2$^+$ & 2.492  & 2.674 & 2.774 \\  
                    & 5/2$^+$ & 0.705  & 0.683  & 0.679 \\                      
            & 1/2$^+$+5/2$^+$ & 2.590 & 2.760 & 2.856                   
\end{tabular}
\end{ruledtabular}
\end{center}
\end{table}
Table \ref{tablo_SMEC1} contains the summary of SMEC  predictions for ANCs in $p$- and $sd$-shell mirror nuclei. To illustrate the impact of continuum coupling, we varied the continuum coupling strength  $V_0$ in the physically relevant range
from 0 (the SM limit) to  $-1.30$ GeV fm$^3$. The resulting ratios $\cal R$ are listed in Table~\ref{tablo_SMEC2}.
\begin{table}[htb]
\begin{center}
\caption{\label{tablo_SMEC2}Ratio $\mathcal{R}$ (\ref{R_ANCs}) calculated in SMEC  for the mirror systems of  Table~\ref{tablo_SMEC1}.
}
\begin{ruledtabular}
\begin{tabular}{c|cccc}
Mirror pair & $j^{\pi}$ &  ${\cal R}_{\rm SM}$  & ${\cal R}_{0.65}$  &  ${\cal R}_{1.30}$  \\ \hline  \\ [-1.5ex]
\multirow{3}{*}{$({^7{\rm Be}}/{^7{\rm Li}})_{3/2^-}$} 
                    & 1/2$^-$ & 1.059  & 1.059  & 1.060  \\
                    & 3/2$^-$ & 1.058  & 1.058  & 1.059 \\
            & 1/2$^-$+3/2$^-$ & 1.058  & 1.058  & 1.060 \\[5pt]  
\multirow{3}{*}{$({^8{\rm Be}}/{^8{\rm Li}})_{2^+}$} 
                    & 1/2$^-$ & 1.055  & 0.974  & 0.584  \\
                    & 3/2$^-$ & 1.048  & 1.048  & 1.047  \\
            & 1/2$^-$+3/2$^-$ & 1.049  & 1.045  & 1.034  \\[5pt] 
\multirow{3}{*}{$({^{12}{\rm N}}/{^{12}{\rm B}})_{1^+}$} 
                    & 1/2$^-$ & 1.333  & 1.333  & 1.331  \\
                    & 3/2$^-$ & 1.323  & 1.333  & 1.347  \\
            & 1/2$^-$+3/2$^-$ & 1.331  & 1.333  & 1.334  \\[5pt]             
\multirow{2}{*}{$(^{17}{\rm F}/{^{17}{\rm O}})_{j^\pi}$} 
                    & 5/2$^+$ & 1.216  & 1.215 & 1.213  \\
                    & 1/2$^+$ & 750.7  & 751.4  & 751.3 \\[5pt]  
\multirow{3}{*}{$({^{18}{\rm Ne}}/{^{18}{\rm O}})_{2_1^+}$} 
                    & 1/2$^+$ & 2.580  & 2.719  & 3.153  \\
                    & 5/2$^+$ & 1.336  & 1.295  & 1.160  \\
            & 1/2$^+$+5/2$^+$ & 2.302  & 2.386  & 2.670  \\[5pt]     
\multirow{3}{*}{$({^{18}{\rm Ne}}/{^{18}{\rm O}})_{2_2^+}$} 
                    & 1/2$^+$ & 28.74  & 27.10  & 19.85  \\
                    & 5/2$^+$ & 5.438  & 7.099  & 11.46  \\
            & 1/2$^+$+5/2$^+$ & 27.02  & 25.88  & 19.37  
\end{tabular}
\end{ruledtabular}
\end{center}
\end{table}

It is seen that the dependence of ANCs and $\cal R$
on the continuum coupling is usually  very weak.
Indeed, in most  considered cases the effect of the continuum coupling on
ANCs does not exceed a few percent, and it is even smaller --a few per mil -- for $\cal R$.
It is instructive to compare ANCs of Tables~\ref{tablo_fe} and \ref{tablo_SMEC1},
and the ratios $\cal R$ of  Tables~\ref{tablo_R} and \ref{tablo_SMEC2}
for the mirror pairs $({^7{\rm Be}}/{^7{\rm Li}})_{3/2^-}$, $({^8{\rm Be}}/{^8{\rm Li}})_{2^+}$, and 
$(^{17}{\rm F}/{^{17}{\rm O}})_{j^\pi}$.
The GSM and SMEC results are extremely consistent when it comes to the the total ANCs (\ref{a4}) and their ratios. 

\begin{figure}[hbt]
\center
\includegraphics[width=0.48\textwidth]{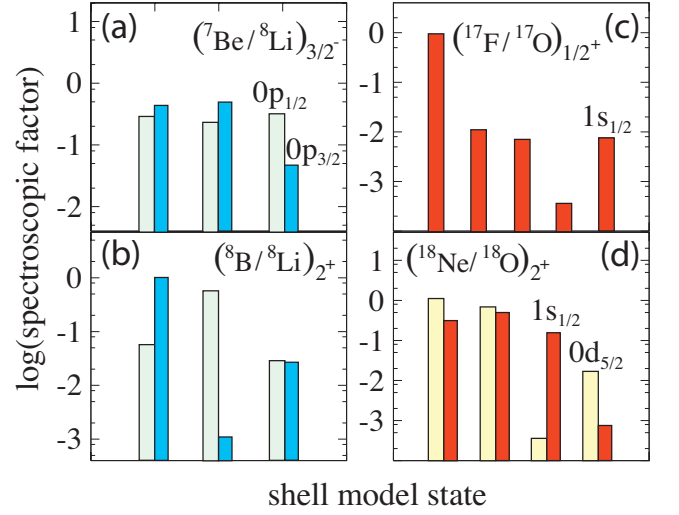}
\caption{\label{fig21x}(Color online) Distribution of spectroscopic strength for SM states coupled to the same decay channel in mirror pairs: $({^7{\rm Be}}/{^7{\rm Li}})_{3/2^-}$ (a); $({^8{\rm B}}/{^8{\rm Li}})_{2^+}$ (b); $(^{17}{\rm F}/{^{17}{\rm O}})_{1/2^+}$ (c); and $(^{18}{\rm Ne}/{^{18}{\rm O}})_{2^+}$ (d).}
\end{figure}

As will be shown below, the effect of the continuum mixing depends on the distribution of spectroscopic factors in SM states coupled to the decay channel. This distribution is shown in Fig.~\ref{fig21x} for selected examples discussed in this section. 

\begin{figure}[hbt]
\center
\includegraphics[width=0.35\textwidth]{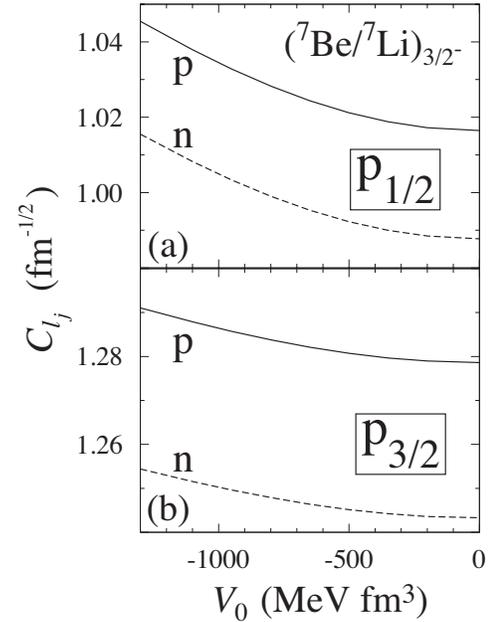}
\caption{\label{fig9}ANCs for mirror nuclei $^7$Be$_{3/2^-}$ (solid line) and $^7$Li$_{3/2^-}$ (dashed line)  calculated in SMEC as a function of the continuum-coupling strength $V_0$. Since the target nucleus $^6$Li has  $J^{\pi}=1^+$, two partial waves are possible:  $p_{1/2}$ (a) and $p_{3/2}$ (b). }
\end{figure}

A typical example, illustrated   in Fig.~\ref{fig9},  shows the variation of ANCs with   $V_0$
for  the mirror pair $({^7{\rm Be}}/{^7{\rm Li}})_{3/2^-}$
and the mirror capture reactions: $^6{\rm Li}_{1^+}+{\rm p}\rightarrow {^7{\rm Be}}_{3/2^-}+\gamma$ and $^6{\rm Li}_{1^+}+{\rm n}\rightarrow {^7{\rm Li}}_{3/2^-}+\gamma$. 
The change of $C_{n\ell_j}$ with $V_0$ is due to the mixing of different $3/2^-$ SM states caused by the continuum coupling. This external mixing of SM states changes the spectroscopic amplitudes 
$S^{1/2}_{p_{1/2}}$ and $S^{1/2}_{p_{3/2}}$ in $C_{p_{1/2}}$ and $C_{p_{3/2}}$ ANCs, respectively. As seen in Fig.~\ref{fig21x}(a), the spectroscopic factors $S_{0p_{1/2}}$ and $S_{0p_{3/2}}$ in $J^{\pi}=3/2_i^-$ states  ($i=1,2,3)$ have all comparable values. The continuum mixing in this case is small,  on the order of 2\% (see Table \ref{tablo_SMEC1}).

Table \ref{tablo_SMEC2} shows that in spite of different proton and neutron  separation energies in the mirror pair $({^7{\rm Be}}/{^7{\rm Li}})_{3/2^-}$, 
the effect of the continuum coupling on the ratio $\cal R$ 
for the 1/2$^-$, 3/2$^-$ partial waves, and the squared norm (\ref{a4}) 
is exceedingly small.

We now consider the mirror pair
$({^8{\rm B}}/{^8{\rm Li}})_{2^+}$ -- involving a proton halo  $^8$B  -- and the mirror reactions: $^7{\rm Be}_{3/2^-}+{\rm p}\rightarrow {^8{\rm B}}_{2^+}+\gamma$ and $^7{\rm Li}_{3/2^-}+{\rm n}\rightarrow {^8{\rm Li}}_{2^+}+\gamma$.  The target nuclei  have $J^{\pi}=3/2^-$; hence,  they can be coupled to the  final $2^+$ state through  $p_{1/2}$ or $p_{3/2}$ waves. 
\begin{figure}[hbt]
\center
\includegraphics[width=0.35\textwidth]{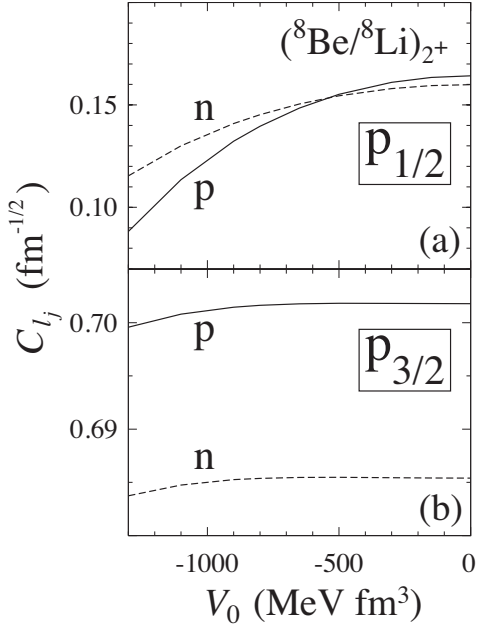}
\caption{\label{fig11}Similar as in Fig.~\ref{fig9} but for mirror nuclei $^8$B$_{2^+}$ (solid line) and $^8$Li$_{2^+}$ (dashed line). Since the target $A=7$ nucleus has  $J^{\pi}=3/2^-$, two partial waves are possible:  $p_{1/2}$ (a) and $p_{3/2}$ (b). }
\end{figure}
Figure~\ref{fig11} shows the corresponding  ANCs. The effect of the continuum coupling is rather important for a small component $p_{1/2}$ and practically negligible in  $p_{3/2}$. Notice also a rather strong --  and different --  dependence of $C_{p_{1/2}}$ on $V_0$ in mirror systems. 
This different response to the continuum-coupling can be traced back  to a different  distribution of SM spectroscopic factors in the three  lowest $2^+$ states; see Fig. \ref{fig21x}(b). 
As the spectroscopic factor $S_{0p_{3/2}}$  of the $2_1^+$ state is close to 1, the state $2_1^+$ is aligned with the decay channel already at $V_0$=0 and no further redistribution of spectroscopic strength is possible through the continuum coupling. 
The situation is different for $S_{0p_{1/2}}$. In this case, the second $2^+$ state has the largest spectroscopic factor and the external mixing leads to a  redistribution of spectroscopic strength; hence, a  change in ANC.
The ratio of ANCs for the mirror pair $({^8{\rm B}}/{^8{\rm Li}})_{2^+}$ is shown in Table \ref{tablo_SMEC2}. The variation of $\cal R$ with $V_0$ is of the order of 1 percent,  and practically the whole effect is due to  the $p_{1/2}$ wave.

As a third example, relevant in the context of GSM analysis, we shall consider the mirror pair $(^{17}{\rm F}/{^{17}{\rm O}})$ in  $J^{\pi}=5/2^+$ ground state  and in the first excited proton halo state $1/2^+$. 
In the  $J^{\pi}=5/2^+$ ground state,  both $^{17}$F and  $^{17}$O couple to $^{16}$O through  the $d_{5/2}$ wave. The $0d_{5/2}$ spectroscopic strength is practically localized in the lowest ${5/2}^+$ state, i.e., the ground state is aligned with a decay channel. The same is true for the  excited $1/2^+$  state,
which practically exhausts the $1s_{1/2}$ spectroscopic strength, see Fig. \ref{fig21x}(c). Consequently, 
as seen in Tables  \ref{tablo_SMEC1} and \ref{tablo_SMEC2}, the continuum coupling is negligible in the $(^{17}{\rm F}/{^{17}{\rm O}})$ case.

In the following, we shall discuss the mirror pair
$({^{18}{\rm Ne}}/{^{18}{\rm O}})$ in the two lowest $2^+$ states, where  the continuum coupling impacts ANCs significantly.
\begin{figure}[hbt]
\center
\includegraphics[width=0.48\textwidth]{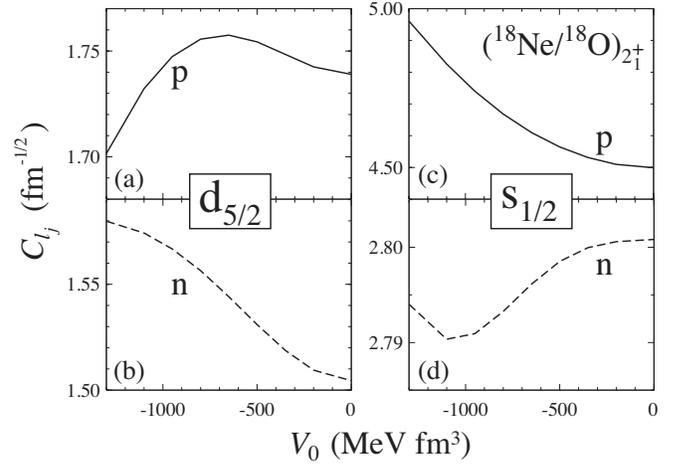}
\caption{\label{fig18}Similar as in Fig.~\ref{fig9} but for mirror nuclei $^{18}$Ne$_{2_1^+}$ (solid line) and $^{18}$O$_{2_1^+}$ (dashed line). Since the target $A=17$ nucleus has  $J^{\pi}=5/2^+$, two partial waves are possible:  $d_{5/2}$ (left panels) and $s_{1/2}$ (right panels).}
\end{figure}
In this case, the corresponding mirror reactions are: $^{17}{\rm F}_{5/2}^+ +{\rm p}\rightarrow {^{18}{\rm Ne}}_{2_i}^+ +\gamma$ and $^{17}{\rm O}_{5/2}^+ +{\rm n}\rightarrow {^{18}{\rm O}}_{2_i}^+ +\gamma$, where $i=1,2$. 
In $^{18}$O, both $2^+$ states are well bound, whereas  in $^{18}$Ne the state $2_2^+$  is close to the proton threshold. It has been shown \cite{cha06} that this state aligns strongly with the decay channel due to the continuum mixing of different $2^+$ SM states.

Figure~\ref{fig18}  shows the mirror ANCs for  $d_{5/2}$ and $s_{1/2}$ partial waves.  It is interesting to notice that with increasing continuum coupling, $C_{d_{5/2}}$ first increases and then strongly decreases in $^{18}$Ne, whereas it steadily increases  in $^{18}$O. As seen in Table \ref{tablo_SMEC1}, the overall  variations in $C_{d_{5/2}}$ are  $\sim8$\%. 
Even stronger variations with $V_0$ are seen for $C_{s_{1/2}}$. In the studied range of $V_0$ values, $C_{s_{1/2}}$ changes by almost $\sim10$\% in $^{18}$Ne  while it varies by $\sim1$\% in $^{18}$O. A different behavior of $C_{d_{5/2}}$ and $C_{s_{1/2}}$ results in a particularly strong variation ($\sim15$\%) of $\cal R$ for the $2_1^+$ state; see Table \ref{tablo_SMEC2}. This behavior can be attributed to the distribution of spectroscopic strength $0d_{5/2}$ ($1s_{1/2}$), which is primarily concentrated in the two (three)  lowest $2^+$ SM states.

\begin{figure}[hbt]
\center
\includegraphics[width=0.48\textwidth]{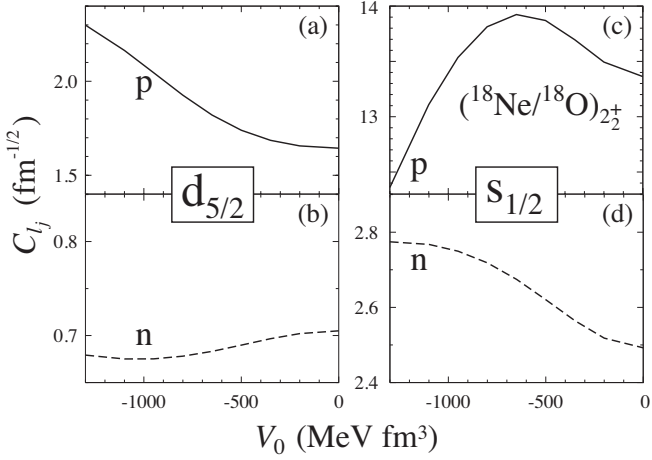}
\caption{\label{fig21}Similar as in Fig.~\ref{fig18} but for $J^\pi=2_2^+$ states in mirror nuclei.}
\end{figure}
Figure~\ref{fig21} shows the ANCs for  $d_{5/2}$ and $s_{1/2}$ partial waves for the  $2_2^+$ state in $^{18}$Ne and $^{18}$O. Also in this case, 
ANCs are strongly affected by the continuum coupling: the ratio $\cal R$ changes by almost 50\% in the considered range of $V_0$, as seen in  Table \ref{tablo_SMEC2}. The distribution of spectroscopic strength $0d_{5/2}$ ($1s_{1/2}$) in this case
is concentrated in the two (three)  lowest $2^+$ SM states; see Fig. \ref{fig21x}(d).

The realistic examples of SMEC calculations of ANCs presented in this section demonstrate that the  distribution of spectroscopic strength over  an ensemble of $J^{\pi}$ SM states is crucial for determining the continuum coupling effect on ANCs. If the spectroscopic strength is  strongly localized in  one  state, as in  $^{17}$F and $^{17}$O, or very broadly distributed, as in $^{7}$Be and $^{7}$Li, then the corresponding ANC is fairly  insensitive to the continuum coupling.
On the other hand, if the spectroscopic strength is concentrated in several close-lying SM states, like in $^{18}$Ne and $^{18}$O,  both mirror ANCs and their ratios may  strongly depend on the coupling to the continuum -- in particular if the state of interest lies close to the particle-emission threshold. 

The  distribution of spectroscopic strength  strongly depends on the effective nucleon-nucleon interaction. In this sense, the quantitative effect of the continuum coupling on ANCs is strongly interaction-dependent. One should keep this in mind when making predictions about mirror reaction cross-sections. For  a given model space and SM interaction, the relative importance of the continuum coupling on mirror ANCs can be {\it a priori} assessed by calculating spectroscopic amplitudes and their distribution in a standard SM. However, the effects of continuum coupling cannot be 
considered in an isolation from the optimization of  the SM interaction to  spectroscopic and reaction observables
within a unified framework. By doing so, an inherent arbitrariness associated with predictions of  ANCs can be reduced.

\subsection{CSM description of ANCs for unbound states}
\label{sec:ANC_unbound}

The definition of ANCs via Eqs. (\ref{eq3},\ref{eq4}) is no longer appropriate
for negative separation energies, i.e., when the state of a nucleus $a$ ($A$-particle system) is
unbound with respect to the nucleus $b$ ($A-1$-particle system). Indeed, in this case  $\kappa$ becomes complex and 
the Whittaker function becomes complex as well. The imaginary part of the Whittaker function is not vanishing even at the limit of vanishing width and the associated ANCs are complex.

\subsubsection{ANC of a complex-energy state and its relation to the particle width}

A suitable definition of the ANC for positive separation energies involves the outgoing Coulomb wave
function $H^+_{\ell,\eta}(k r)$, where $k=(-2\mu S_a/\hbar^2)^{1/2}$ and $\eta=Z_bZ_ce^2\mu/\hbar^2 k$:
\begin{eqnarray}
I^a_{bc;\ell j}(r) \sim \frac{1}{r} C_{\ell j} H^+_{\ell,\eta}(k r).
\label{eq1_unbound}
\end{eqnarray}
Note that $k$ and $\eta$ are complex \cite{my07}, as the state in $A$-particle systems is 
unbound. At the limit of vanishing width, $k$, $I^a_{bc;\ell j}$ and $H^+_{\ell,\eta}(k r)$ become real, so that
 $C_{\ell j}$ given  by Eq.~(\ref{eq1_unbound}) becomes real as well.

For narrow resonances,  ANCs can be related to  the particle width  \cite{Muk99}. However, the derivation of this relationship in 
 Ref.~\cite{Muk99} relies on the R-matrix theory  --
 not used in the context of GSM --  so it is useful to recall the derivation for the Gamow states.
The overlap function $I^a_{bc;\ell j}(r)$ defined in Eq.~(\ref{eq1})
obeys a Schr\"odinger-like equation, albeit inhomogeneous \cite{Pinkston_Satchler}.
However, separating the full interaction into a one-body term  and a two-body residual interaction,
the source term can be decomposed  into a dominant homogeneous part and a
residual inhomogeneous part, and  the latter can be absorbed into the homogeneous part.
This approximation has been tested successfully in Ref.~\cite{my07} for both bound and unbound states.
Moreover, as only narrow resonant states are involved, we will consider
that the potential entering the equation defining $I^a_{bc;\ell j}(r)$
is real. This simplification breaks down for resonant states
bearing a sizeable width, for which complex potentials must be used \cite{my07},
but is sound for narrow resonances.

Under these assumptions, one can easily derive the  relation between ANC and partial width \cite{Humblet_Rosenfeld}. The function 
$\tilde{I}(r) \equiv  r I^a_{bc;\ell j}(r)$ is a solution of the  Schr\"odinger equation:
\begin{equation}\label{u_eq}
\tilde{I}''(r) = \left( \frac{\ell(\ell+1)}{r^2} + v(r) - k^2 \right) \tilde{I}(r),  
\end{equation}
where $v(r)$ is real and local. 
The boundary conditions defining $\tilde{I}(r)$ are $\tilde{I}(r=0) = 0$ and Eq.~(\ref{eq1_unbound}) at large $r$. The continuity equation for $\tilde{I}(r)$ implies 
that  
\begin{eqnarray}
&&{\tilde{I}}{^*}(r) \tilde{I}'(r) - \tilde{I}'{^{*}}(r) \tilde{I}(r) \nonumber \\
&=& ({k^*}^2 - k^2) \int_0^r |\tilde{I}(s)|^2~ds. \label{W_eq} 
\end{eqnarray}
Taking $r$ in the asymptotic zone in which (\ref{eq1_unbound}) applies,
 noticing that
${k^*}^2 - k^2$ is proportional to the partial width $\Gamma_{\ell j}$,
and utilizing the standard mirror relation for Coulomb wave functions ${H^+_{\ell,\eta}(z)}^* = H^-_{\ell,\eta^*}(z^*)$, 
(both functions obey the same differential equation and behave as $\exp(-i z^* + i\eta^* \ln(2 z^*))$ for $|z| \rightarrow +\infty$),
one obtains:
\begin{eqnarray}
\Gamma_{\ell j} &=& \frac{k H^-_{\ell,\eta^*}(k^* r) {H^+_{\ell,\eta}(k r)}' - k^* {H^-_{\ell,\eta^*}(k^* r)}' H^+_{\ell,\eta}(k r)} {2i \int_0^r |\tilde{I}(s)|^2~ds} \nonumber \\
&\times& \frac{\hbar^2}{\mu} |C_{\ell j}|^2, \label{Gamma_eq1}
\end{eqnarray}
where $\mu$ is the effective mass of the particle.
To get rid of the explicit $r$-dependence in Eq.~(\ref{Gamma_eq1}), further approximations are necessary  \cite{Barmore, *Kru04}.
Neglecting $\Im(k)$ in the Coulomb wave functions of the numerator of Eq.~(\ref{Gamma_eq1})
implies that their Wronskian becomes equal to $2i\Re(k)$. Moreover, as $\tilde{I}(r)$ has a
quasi-bound state character, it decreases exponentially along the real $r$-axis (unless $r$ becomes extremely large, which we do not consider here), 
so that the integral in the denominator is almost equal to one when $r$ is chosen in the asymptotic region. Under these assumptions, valid for narrow resonances,
Eq.~(\ref{Gamma_eq1})  simplifies to:
\begin{equation}\label{narrow_width_formula}
\Gamma_{\ell j} = \frac{\hbar^2}{\mu} |C_{\ell j}|^2 \Re(k),
\end{equation}
which is the same expression  as that obtained in  Ref.~\cite{Muk99},
even though approximations and boundary conditions are different in the real-energy R-matrix approach and complex-energy Gamow-state formalism \cite{Barmore, *Kru04}.

Expressing the  total width $\Gamma$ in terms of  the sum of partial widths $\Gamma_{\ell j}$
gives  total ANC constant $C$ (\ref{Cnorm}):
\begin{equation}\label{total_C_formula}
C = \sqrt{\sum_{\ell j} |C_{\ell j}|^2}=\sqrt{\Gamma \frac{\mu}{\hbar^2 \Re(k)}}. 
\end{equation}

\subsubsection{CSM description of ANCs for unbound states}
\label{sec:GSMu}

\begin{figure}[hbt]
\center
\includegraphics[width=0.35\textwidth]{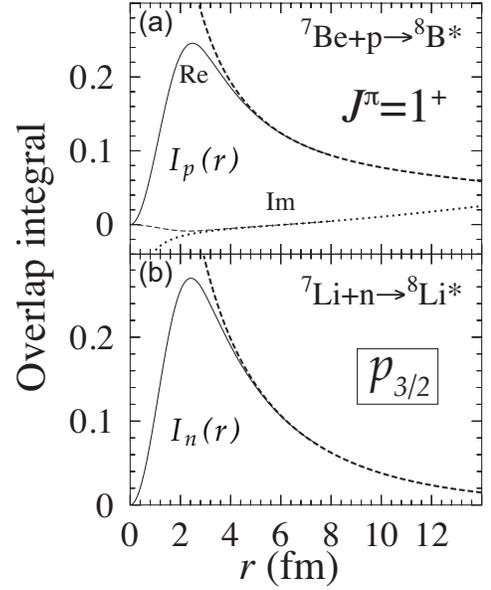}
\caption{\label{Ov3-GSM}Similar as in Fig.~\ref{Ov1-GSM} but for 
 $^7{\rm Be}_{3/2^-}+p\rightarrow {^8{\rm B}}_{1^+}$ (a) and $^7{\rm Li}_{3/2^-}+n\rightarrow {^8{\rm Li}}_{1^+}$ (b).}
\end{figure}

\begin{figure}[hbt]
\center
\includegraphics[width=0.35\textwidth]{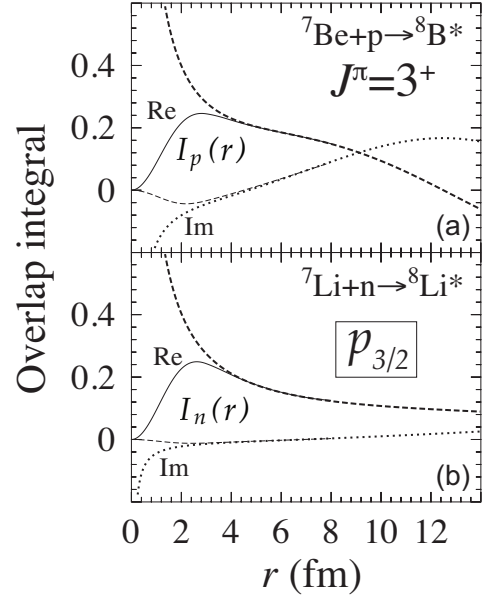}
\caption{\label{Ov4-GSM}Similar as in Fig.~\ref{Ov3-GSM} but for the $J^\pi=3_1^+$  resonance. In $^8$B this state is a broad one-proton resonance which results in a large imaginary part of radial overlap integral.
}
\end{figure}

Figure~\ref{Ov3-GSM} compares radial overlap integrals  (\ref{eqex}) for the excited $J^{\pi}=1_1^+$  states of $^8$B and $^8$Li in the  $p_{3/2}$  channel. 
Since the   $1_1^+$ state of $^8$B is a narrow one-proton resonance, the proton overlap integral acquires   a small imaginary part.
Figure \ref{Ov4-GSM} shows the radial overlap integrals for a $3_1^+$ 
broad resonance in $^8$B and a narrow mirror resonance in $^8$Li. The tails of  real and imaginary parts of  radial overlap integrals are fitted with the outgoing Coulomb wave functions of a complex argument $k$. 
Table~\ref{tablo_de} displays 
proton and neutron ANCs and their ratio $\mathcal{R}$
for the first excited  $J_1^{\pi}=1_1^+$ state in $^8$B  and $^8$Li.
\begin{table}[htb]
\begin{center}
\caption{\label{tablo_de}Proton and neutron ANCs,    $\mathcal{R}$,  and  $\mathcal{R}_\Gamma$ (\ref{a5_2}) calculated in GSM
for the    $J^{\pi}=1_1^+$  excited state in $^8$B (one-proton resonance) and $^8$Li. See text for details.
}
\begin{ruledtabular}
\begin{tabular}{c|cccc}
 $j^{\pi}$ &   $C_p$   &  $C_n$  & ${\cal R}_{\rm GSM}$  & ${\cal R}_\Gamma$ \\ \hline  \\ [-1.5ex]
 1/2$^-$           & 0.0322$-i$0.00138   & 0.1379 & 0.0545 & 0.0021 \\ 
 3/2$^-$           & 0.0442$-i$0.00273   & 0.2090 & 0.0449 & 0.0018 \\
1/2$^-$+3/2$^-$    & 0.0547              & 0.2504 & 0.0478 & 0.0019      
\end{tabular}
\end{ruledtabular}
\end{center}
\end{table}

\subsubsection{SMEC description of ANCs for unbound states}
\label{sec:SMECu}

To relate ANCs in bound-unbound mirror pairs
and extract the proton decay
width  from the neutron ANC in the bound mirror state,
Timofeyuk {\it et al.}  \cite{timo03, *Timdes05a} introduced a quantity:
\begin{equation}
\mathcal{R}_\Gamma =\frac{1}{\hbar\,c}\,\frac{\Gamma_p}{|C_n|^2},
\label{a5}
\end{equation}
where $|C_n|$ is given by  Eq.~(\ref{a4}) and $\Gamma_p$ is the total width:
\begin{eqnarray}
\Gamma_p=\sum_{\ell,j} \Gamma_{\ell j} \left|S_{\ell j}\right|,
\label{a6}
\end{eqnarray}
where $\Gamma_{\ell j}$ is the partial proton width of a Gamow state with an 
quantum numbers $\ell$ and $j$. The modulus of a spectroscopic factor $S_{\ell j}$ is taken
in order to ensure that $\Gamma_p$ remains positive after the coupling to the particle continuum. 
For narrow resonances, using Eq.~(\ref{narrow_width_formula}), one can
express $\mathcal{R}_\Gamma$ in terms of proton and neutron ANCs:
\begin{eqnarray}
\mathcal{R}_\Gamma = |C_p/C_n|^2 ~ \Re(k_p) ~ \frac{\hbar c}{\mu_p c^2}. 
\label{a5_2}
\end{eqnarray}

\begin{figure}[hbt]
\center
\includegraphics[width=0.35\textwidth]{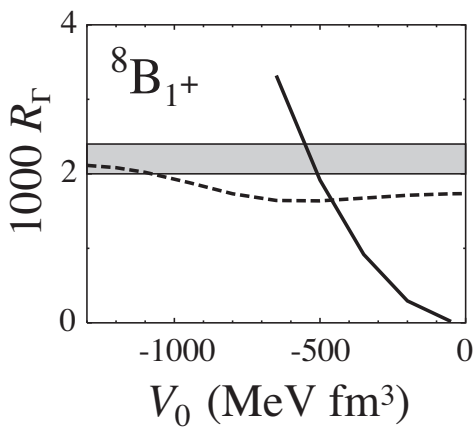}
\caption{\label{fig25}The dependence of $\mathcal{R}_\Gamma$ (\ref{a5}) in SMEC on $V_0$ in mirror systems $^{8}$B$(1_1^+)$ (proton resonance) and $^{8}$Li$(1_1^+)$ (bound state). $\mathcal{R}_\Gamma$  using  Eq.~(\ref{a5}) 
(solid line) is compared to that using   expression  (\ref{a6}) for the proton decay width (dashed line). The shaded area shows the range of  $\mathcal{R}_\Gamma$ extracted in Ref.~\cite{timo03} using  experimental  values of $\Gamma_p$ \cite{ajz88} and $C_p$ \cite{tra03}.
}
\end{figure}

As an example, let us consider the previously discussed case of the  proton  $1^+$ resonance in $^8$B and its bound mirror analog in $^8$Li. 
Figure~\ref{fig25} shows the continuum coupling strength dependence of $\mathcal{R}_\Gamma$ computed in SMEC. The SMEC prediction is compared with   the approximate formula (\ref{a6}) using  the spectroscopic factors obtained in SMEC. The shaded region correspond to the estimate of Ref.~\cite{timo03} based on experimental data for mirror states in $^8$B and $^8$Li.
The SM prediction is $\mathcal{R}_\Gamma^{\rm SM}=1.73\times 10^{-3}$, whereas the experimental value extracted in Ref.~\cite{timo03} is $(2.2\pm0.2)\times 10^{-3}$. Results of SMEC obtained using Eq.~(\ref{a6}) for the proton  width become compatible with the experimental results for $V_0<-1000$~MeV~fm$^3$. The GSM prediction  $\mathcal{R}_\Gamma=1.88 \times 10^{-3}$ given  in Table~\ref{tablo_de} is fairly close to experiment
and to SMEC values.
The SMEC values obtained  with Eq.~(\ref{a5}) show a strong dependence on $V_0$. This estimate of $\mathcal{R}_\Gamma$ is valid in the range of  $0<\Gamma_p<\Gamma^{(sp)}$, where $\Gamma^{(sp)}$ is the  s.p.~width, which in this case is 64.9 keV  and 62.6 keV for $p_{3/2}$ and $p_{1/2}$, respectively.

\begin{figure}[hbt]
\center
\includegraphics[width=0.35\textwidth]{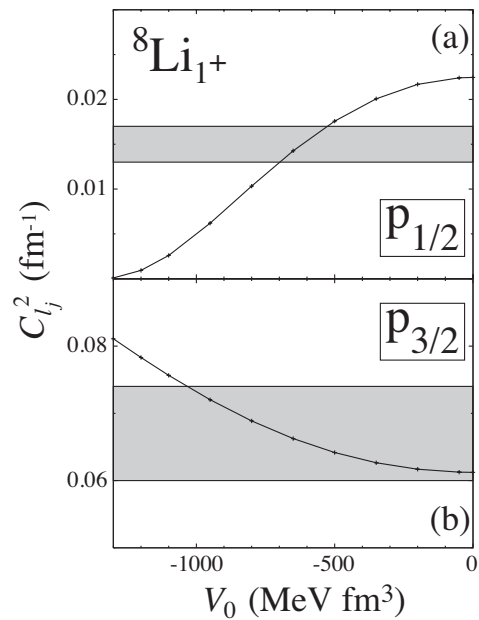}
\caption{\label{fig26}Squared ANCs of the $p_{1/2}$ (a) and $p_{3/2}$ (b) neutron contributions to  $^{8}$Li$_{1_1^+}$. Shaded area mark  experimental uncertainties on these quantities \cite{timo03}.
}
\end{figure}

The 
$p_{1/2}$ and $p_{3/2}$ neutron contributions to the ANC of $^{8}$Li$_{1_1^+}$ are shown in Fig.~\ref{fig26}. 
One  can see that $p_{1/2}$ and $p_{3/2}$ contributions  have different dependence on $V_0$ which, in turn, reduces  variations in $\mathcal{R}_\Gamma$. In general, for a physical range of $V_0$ in this mass region, $-500$~MeV~fm$^3>V_0>-1200$~MeV~fm$^3$, SMEC  agrees somewhat better with  experiment than  SM.

\section{Outlook}
\label{sec:Concl}

In the first part of the paper, we discussed
the basic properties of ANCs and SPANCs. We broadly classified the behavior of SPANCS for charged and neutral particles in terms of the Sommerfeld parameter
$\eta$ and the wave number $\kappa$, respectively, as well as  the orbital angular momentum. The  extreme regimes of SPANC for charge particles can be characterized
by the complex turning point $z_t$ of the outgoing Coulomb wave function.
We also discussed the near-threshold  behavior of ANCs.

Based on the argument using the charge symmetry of the nuclear force,  a simple relation (\ref{Ras}) between proton and neutron  ANCs in mirror pairs has been  proposed \cite{timo03, *Timdes05a}.  The estimate ${\cal R}_0$ is very useful as it relates cross-sections of low-energy direct and resonance proton capture reactions, which  are difficult or impossible to measure, with neutron ANCs obtained in reactions with stable beams. In the second part of this study, the link between mirror ANCs through relation (\ref{Ras})  has been verified in our CSM calculations for different physical situations of the coupling to the scattering continuum and for various many-body  states. It has been found that the key factor in determination of ANCs and the  mirror ratio ${\cal R}$, as well as the sensitivity of ANCs to continuum coupling,   is the distribution of spectroscopic strength that is both model- and interaction-dependent.  For example, relative differences of ${\cal R}$ in GSM and VMC \cite{Nol11} can be as large as ~30\%, and the continuum coupling can change $\cal R$ by up to 50\% in exceptional cases. Also,  differences with respect to ${\cal R}_0$  can be non-negligible. In this sense, ANCs and their mirror ratios are interaction-dependent. 

It has been found  that the quantitative effect of the continuum coupling on ANCs and their ratios is  minor if the spectroscopic strength is either localized in a single SM state or broadly distributed. This property is
independent on binding energies of mirror states. On the other hand, if the spectroscopic strength is concentrated in several SM states, their coupling via the continuum space may result in  a significant rearrangement of the spectroscopic strength; hence, appreciable variations of ANCs with respect to SM predictions. This effect is particularly strong for near-threshold states that align with the decay channel. 
Since these special cases can be {\it a priori} identified in standard SM calculations of the spectroscopic strength distribution, the qualitative effect of the continuum coupling on SM results for ANCs  and their ratios can easily be assessed without resorting to  sophisticated CSM calculations, which ultimately provide the quantitative answer.

Finally, let us state that uncertainties due to the  model dependence 
can be significantly reduced if the effective SM interaction is optimized to both  spectroscopic and reaction observables within a unified SCM framework. Work along these lines is in progress.

\bigskip
\begin{acknowledgments}
Useful discussions with Filomena Nunes and Luke Titus are gratefully acknowledged.
This work was supported  by the Office of
Nuclear Physics,  U.S. Department of Energy under Contract Nos.
DE-FG02-96ER40963 (University of Tennessee) and DE-FG02-10ER41700 (French-U.S. Theory Institute for Physics with Exotic Nuclei); by MSWiN Grant No.~N~N202~033837; by the CICYT-IN2P3 cooperation; and  by the Academy of Finland and University of Jyv\"äskyl\"ä within the FIDIPRO programme.

\end{acknowledgments}


\bibliographystyle{apsrev}
\bibliography{ANC}

\end{document}